\documentclass[11pt]{article}
\usepackage{cite}
\usepackage{amsmath,amsfonts,amssymb}
\usepackage[small,bf,hang]{caption}
\usepackage[dvips]{graphicx,epsfig}
\usepackage{slashed}
\usepackage{epsfig}
\usepackage{graphicx}
\usepackage{epstopdf}
\usepackage[vcentermath]{youngtab}

\def\hybrid{
        \topmargin -20pt
        \oddsidemargin 0pt
        \headheight 0pt \headsep 0pt
        \textwidth 6.25in 
        \textheight 9.5in 
        \marginparwidth .875in
        \parskip 5pt plus 1pt \jot = 1.5ex}

\hybrid
\usepackage{tikz}

 \csname
@addtoreset\endcsname{equation}{section}

\usepackage{color}
\definecolor{red}{rgb}{1,0,0}
\definecolor{lred}{rgb}{0.3,0,0}
\definecolor{green}{rgb}{0,0.6,0}
\definecolor{blue}{rgb}{0,0,1}
\definecolor{violet}{rgb}{0.8,0,0.8}
\definecolor{amber}{rgb}{1.0, 0.75, 0.0}
\definecolor{yellow}{rgb}{1.0, 1.0, 0.0}
\definecolor{applegreen}{rgb}{0.55, 0.71, 0.0}
\definecolor{cadmiumgreen}{rgb}{0.0, 0.42, 0.24}
\definecolor{ballblue}{rgb}{0.13, 0.67, 0.8}
\definecolor{caribbeangreen}{rgb}{0.0, 0.8, 0.6}
\definecolor{bluemunsell}{rgb}{0.0, 0.5, 0.69}
\definecolor{brightpink}{rgb}{1.0, 0.0, 0.5}

\def\moth{\mathsurround=0pt}
\newdimen\zo \zo=0pt

\def\tick{\leaders\hrule height 0.5ex depth 0pt \hskip 0.5pt}
\def\upboxfill{$\moth \setbox\zo\hbox{\tick}%
  \hskip 3pt\hbox to 0pt{$\tick$\hss}\hrulefill \hbox to 7.5pt{$\tick$\hss}$}

\def\dtick{\leaders\hrule height .34pt depth 0.5ex \hskip 0.5pt}
\def\downboxfill{$\moth \setbox\zo\hbox{\dtick}%
  \hskip 2pt\hbox to 0pt{$\dtick$\hss}\hrulefill \hbox to 2pt{$\dtick$\hss}$}

\def\bec{\begin{center}}
\def\ec{\end{center}}

\newcommand{\eq}[1]{(\ref{#1})}

\def\be{\begin{equation}}
\def\ee{\end{equation}}
\def\bea{\begin{eqnarray}}
\def\eea{\end{eqnarray}}
\def\ba{\begin{array}}
\def\ea{\end{array}}

\begin{document}

\begin{titlepage}
	\rightline{}
	\rightline{August  2023}
	\rightline{HU-EP-23/46-RTG} 
	\rightline{MIT-CTP/5595} 
	\begin{center}
		\vskip 1.1cm

  {\Large 
		 \bf{On black hole singularity resolution  in $D=2$   }  \\[1.0ex]
		 {via  duality-invariant $\alpha'$ corrections    }}

		\vskip 1.5cm

{\large\bf {Tomas Codina$^\dag$, Olaf Hohm$^\dag$ and Barton Zwiebach$^*$}}
		\vskip 1cm
		
		$^\dag$ {\it   Institute for Physics, Humboldt University Berlin,\\
			Zum Gro\ss en Windkanal 6, D-12489 Berlin, Germany}\\
		
		\vskip .3cm
		
		$^*$ {\it   Center for Theoretical Physics, \\
		Massachusetts Institute of Technology, \\
		Cambridge MA 02139, USA}\\
		\vskip .1cm
		
		\vskip .4cm

		tomas.codina@physik.hu-berlin.de, ohohm@physik.hu-berlin.de, zwiebach@mit.edu

		\vskip 1.9cm
		\end{center}
	
\begin{quote} 	
		
\centerline{\bf Abstract} 

\medskip

Starting with the two-derivative limit of $D=2$ string theory,
we explore the space of T-duality invariant $\alpha'$ corrections, 
a space that contains
a point representing the fully $\alpha'$-corrected classical string theory. 
Using a parametrization introduced by Gasperini and Veneziano 
we obtain black hole solutions in this theory space. 
We prove that the dual of a solution with a regular horizon must
have a curvature singularity. 
We find regions in the theory space where the black hole is deformed while preserving the horizon and the singularity, and regions where no black hole appears
to exist.  
Furthermore,  we find subregions in this  
theory space, probably not containing string theory, 
in which the black hole geometry exhibits
a horizon leading to an interior that, having no singularity in the metric, curvature, or dilaton, is a regular cosmology.

	\end{quote} 
	\vfill
	\setcounter{footnote}{0}
	
	\setcounter{tocdepth}{1} 
	
\end{titlepage}

\tableofcontents

\section{Introduction } 

Black holes are perhaps the most mysterious objects predicted by general relativity. 
They  are characterized 
by an event horizon that forms, for instance,  when the size of a collapsing star falls below its 
Schwarzschild radius. It follows from  the singularity theorems of Penrose that in general relativity the gravitational collapse inevitably leads to a singularity at which the spacetime curvature becomes infinite \cite{Penrose:1964wq}.
Since singularities cannot exist in nature we learn that 
general relativity breaks down close to the black hole singularity. 
In contrast, general relativity should be applicable at the event horizon, since the curvature there might be relatively small. In addition, we have excellent reasons to believe that  
black holes with  an event horizon exist in nature.
 This then leads to the problem of resolving the black hole singularity: finding a new theory 
that replaces general relativity in the appropriate regime so that there are solutions with 
an event horizon but no spacetime singularity in the interior \cite{Bardeen,Trodden:1993dm,Bronnikov:2006fu,Lemos:2011dq,Lan:2023cvz}.   

In this paper we ask if the black hole (BH) 
singularity can be resolved in string theory by means of the higher-derivative $\alpha'$ corrections that are already a feature of classical string theory. 
We consider non-critical string theory in two dimensions (2D). 
Apart from its simplicity this case is particularly promising in that there is an exact worldsheet CFT 
whose target space interpretation is that of a 2D BH
\cite{Witten:1991yr, Mandal:1991tz,Giveon:1991sy}. Thus there should be an exact BH solution to all orders in $\alpha'$. 
We were motivated by the recent result of 
Gasperini and Veneziano~\cite{Gasperini:2023tus} who showed 
that,  remarkably,  
the big-bang singularity may be resolved in the $\alpha'$-complete theory.
Since the BH interior may be visualized as a cosmology, one could
envision that a resolution of the BH singularity is feasible.
Recent approaches trying to remove the BH singularity using the setup developed in~\cite{Hohm:2019jgu, Hohm:2019ccp} 
for cosmological backgrounds were studied in \cite{Wang:2019dcj,Ying:2022xaj, Ying:2023gmd, Ying:2023gnn}. 
An old result in this direction is the $D=2$ black-hole geometry 
of~\cite{Dijkgraaf:1991ba}, which was suggested to be $\alpha'$-exact.   
The maximally-extended space-time geometry and regularity of such 
ansatz 
 was studied in~\cite{Tseytlin:1991xk,Perry:1993ry, Yi:1993gh}
 and an action describing this 
background and its (singular) dual   
was built in~\cite{Grumiller:2005sq}.

This paper is based on our recent results in \cite{Codina:2023fhy}, in which we classify the possible higher-derivative corrections that are 
compatible with an $O(1, 1)$ duality symmetry that string theory is known to possess for such backgrounds to all orders in $\alpha'$ \cite{Meissner:1991zj,Meissner:1991ge,Sen:1991zi,Giveon:1994fu,Meissner:1996sa}. 
In this we generalized the earlier work in  \cite{Hohm:2019jgu, Hohm:2019ccp} on cosmology. 
Specifically, we consider the class of 2D string backgrounds with metrics 
$ds^2 = - m^2 (x) dt^2  + n^2(x) dx^2$ that admit a time-like isometry since the metric 
components are independent of time $t$. The outside region of the  
2D BH is of this form. For any local solution  of this form, in string theory there is also 
the  T-dual solution  obtained by sending $m\rightarrow \frac{1}{m} $,  
while $n$ and the dilaton $\Phi$, 
defined in terms of the standard dilaton via 
$\sqrt{- g} e^{-2 \phi} = n e^{-\Phi}$, are invariant.  
By exploiting all possible perturbative field redefinitions  that are consistent with 
duality we showed, under certain assumptions on the fall-off behavior of the coefficients, 
that the most general duality invariant action is of the form 
\begin{equation}\label{IallINTRO}
I = \int dx \, n\,  e^{-\Phi} \Bigl[ 1
+ (D\Phi)^2 + F(M)\Bigr]\,,
\end{equation} 
where 
$D:=\frac{1}{n}\partial_x$ is a covariant derivative,  
$M:=D \log m$,  and 
\begin{equation}
F(M) \equiv \sum_{i = 0}^{\infty} \epsilon_i M^{2i + 2} = - M^2 + \epsilon_1 M^4 + \dots\,, \quad \epsilon_0 \equiv -1\,. 
\end{equation}
The $\epsilon_i$ are the free coefficients that are not fixed by duality and that must be determined 
by other methods from string theory (see for instance \cite{Codina:2020kvj,Codina:2021cxh,Bonezzi:2021sih}). We currently do not have the techniques to determine them and hence the function $F(M)$ to all orders. Nevertheless, here we explore the question whether 
there are points in the `theory space' of all choices of $F(M)$ 
for which there are  
BH solutions 
that carry an event horizon but are regular everywhere. 

In order to deal efficiently with the equations following from the $\alpha'$-complete action (\ref{IallINTRO}) 
we will use a powerful representation  
recently introduced by Gasperini and Veneziano~\cite{Gasperini:2023tus}, 
to study the resolution of  
the big-bang singularity (for earlier work 
in this direction
see~\cite{Wang:2019kez,Ying:2021xse}). 
In the present black-hole context,   
instead of seeking a solution given by explicit functions $m(x)$ and $\Phi(x)$ (say fixing the gauge $n=1$) 
one seeks a solution of the form $m(f)$, $\Phi(f)$ and $x(f)$, with $f$ a new parameter. 
Remarkably, such a parametrized form can be obtained by 
inverting  
the function $f(M):=F'(M)$ to obtain $M(f)$, which parametrizes the metric variable
$M$ in terms of $f$.
For the two-derivative theory $f(M) = - 2M$ and $M(f) = -\tfrac{1}{2} f$. 
While perturbatively $\alpha'$ corrections build up the function $f(M)$ as a power series,
the inverse function $M(f)$ affords a more direct solution of the equations of motion.  
We are thus led to think of the `theory space' of duality-invariant derivative corrections as one described by the possible choices of the function $M(f)$.
This leads to subtle questions on how to encode a complete solution by such a Gasperini-Veneziano
parametrization, which in general requires the use of several contours 
in `$f$-space.' Each contour represents a patch in coordinate
space, and together they represent the whole space.  
In fact, the equations
of motion imply the existence of branch points and branch cuts in the $f$ plane, 
and the underlying structure of the solution is determined by these features.
The relation $x(f)$ is given in terms of an integral $x = \int^f \cdots$.  
We divide a complete solution into {\em exterior} and {\em interior} regions, a division that has a clear meaning in the context of black hole solutions. We will use $f$ to parameterize the exterior and $\tilde f$ for the interior, to make clear they represent different patches.
 
\begin{figure}[h]
	\centering
\epsfysize=6.8cm
\epsfbox{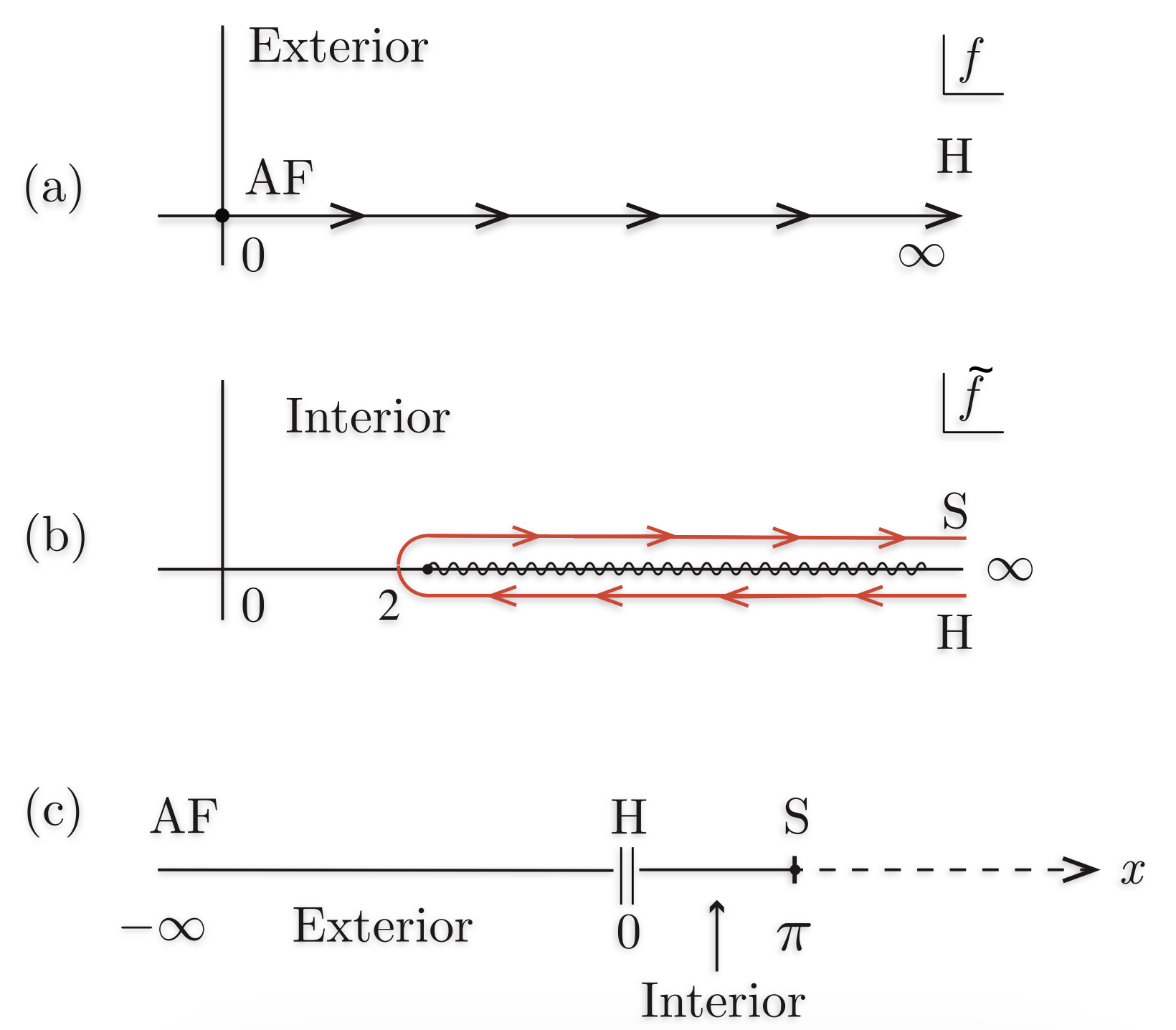}
	\caption{\small 
	The two-derivative black hole. (a) The exterior is produced by a contour $f\in (0, \infty)$ joining the asymtotically flat (AF) region to the horizon (H).  (b)  The interior requires a contour going under and then over the cut, with branch point $\tilde f=2$ and the horizon and singularity at $\tilde f=\infty$, under and over the cut, respectively.
 (c) The $x$-space representation of the black hole. The exterior is in $(-\infty, 0)$ and the interior in $(0, \pi)$.} 
	\label{f1int}
\end{figure}

For the two-derivative BH the {\em exterior} geometry is obtained with an $f$ plane contour going from zero to infinity on the real line (Fig.\,\ref{f1int}(a)).
Here $f=0$ is the asymptotically flat (AF) region and $f=\infty$ is the horizon. 
For the interior, we find that the $\tilde f$ plane has a branch point at $\tilde f=2$ and a cut
running from this point to infinity.  The contour that gives the interior solution begins
under the cut at infinity goes down to the branch point and then returns to infinity
at the top of the cut (Fig.\,\ref{f1int}(b)).  The horizon and the singularity are at $\tilde f=\infty$, below and above the cut, respectively.  The $x$-space picture has the
exterior going from $x=-\infty$ (AF) to $x=0$ (the horizon).  The interior, drawn
on the same picture goes from the horizon at $x=0$ to the singularity at 
$x=\pi$ ((Fig.\,\ref{f1int}(c)).  
It must be emphasized that, in general, different $x$-space coordinate patches are used to 
cover the BH solution. For instance, in the coordinate systems we mainly use for the standard BH,  the $x$ coordinate ranges  
from $-\infty$ to $0$ and then from $0$ to $\pi$, which are two different patches, despite being displayed in the same graph.

It is worth noting that since the roles of space and time are interchanged when passing the event horizon, 
the isometry becomes space-like, and hence the interior of a black hole is actually a cosmology~\cite{Tseytlin:1991xk}.  
For this cosmology the time coordinate is $x$,  
with time $x=0$ corresponding to the event horizon. Since the curvature is finite at the event horizon, 
the cosmology so obtained does not have a big-bang singularity. In contrast, 
after a finite time 
the scale factor becomes infinite, with infinite curvature, 
leading to a `big rip'.

In this paper we show that $T$ duality implies
that the dual of a solution with a regular horizon must have a curvature singularity. 
This result, valid for any solution in the full theory space, is a formalization of the early observation
of Giveon~\cite{Giveon:1991sy} noting this fact in the context of the two-derivative version of the string theory black hole.  It has some implications in our work; in particular, two
options seem available. 
First, solutions may be self-dual in the sense that the full BH (extended) 
geometry is invariant under T-duality.  In such a case, the BH must have a curvature
singularity somewhere if it has a horizon.  This is the case for the two-derivative
black hole solution.  Second, a complete solution with a horizon {\em and} no curvature or dilaton singularity would not be self-dual, but it would still be physically equivalent
(in the sense of string theory) to its $T$-dual, which necessarily would exhibit a
curvature singularity. 
It should be stressed that throughout this paper   
we employ the standard general relativity notion of singularities, which is probed 
by point particles. In string theory, however, spacetime is probed  
by strings, which requires further investigation.

We show that for a large class 
of functions $M(f)$ there are black hole solutions that, just like that of the two-derivative theory,   
are not regular and necessarily carry  a 
singularity. 
More precisely, parametrizing the space of functions as  
$M(f) = - \frac{f}{2} \left(1 + h(f^2)\right)$, we find a number of constraints 
that the function $h$ has to obey in order to obtain solutions with a horizon. 
These conditions thus pose rather stringent constraints on the $\alpha'$ corrections 
for string theory to permit BH solutions with a horizon. Conversely, if one could show that the functions  $h$ arising in string theory do not obey these constraints one could prove that string theory does not permit 
BH solutions.\footnote{The work of S.~Ying in \cite{Ying:2022xaj} 
appears to imply that a  cubic $M(f)$  
would yield a regular BH.  Our analysis of cubic deformations in this
paper, however,  finds no such regular solution. In fact,~\cite{Ying:2023gmd} indicates that this regular solution requires changing the value of the string theory constant term in the action.}

\begin{figure}[h]
	\centering
\epsfysize=6.9cm
\epsfbox{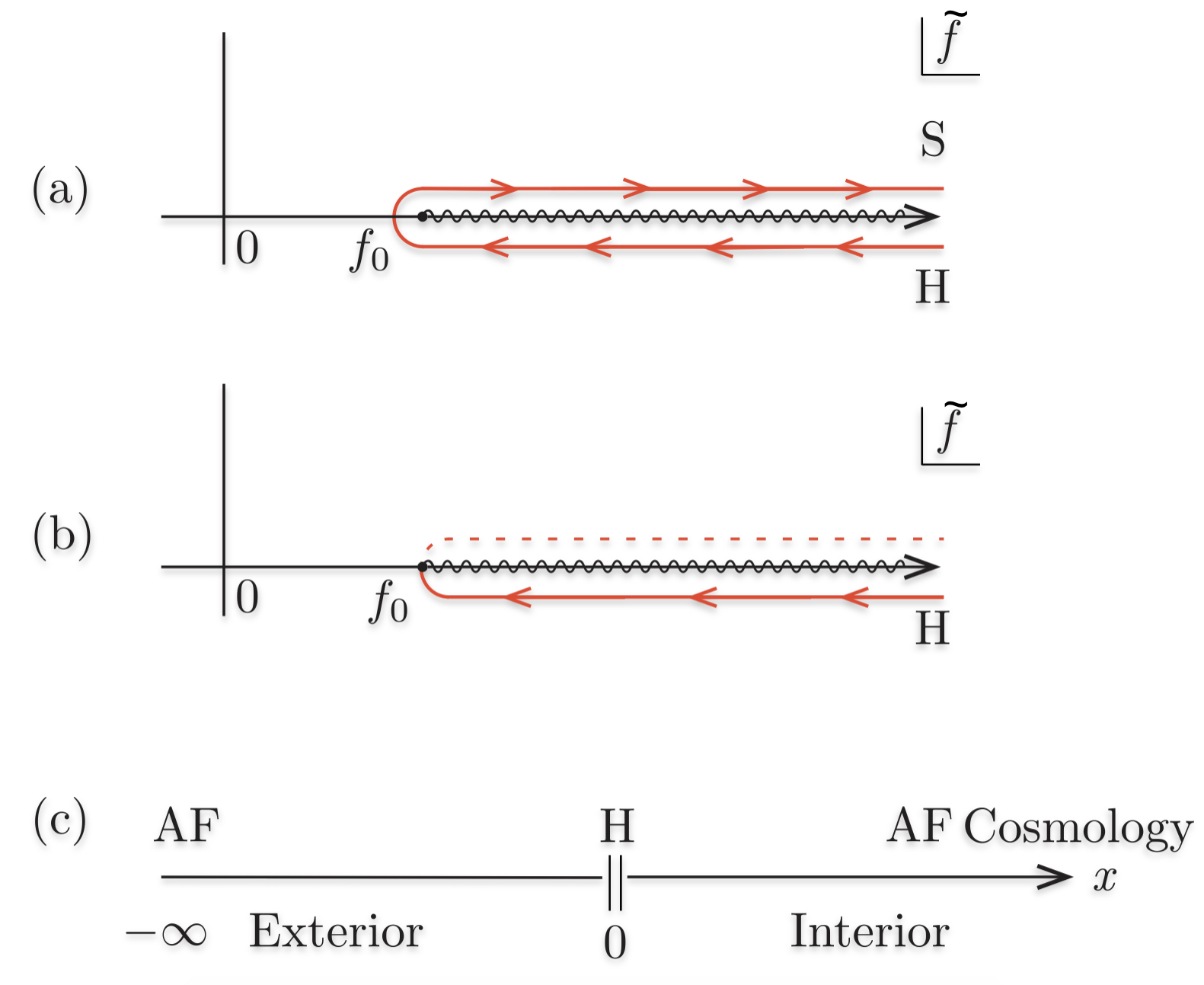}
	\caption{\small  (a) The interior of a deformed black hole with a full contour
	containing a horizon H 
	and a singularity S. (b)  The interior of the regularized black hole, without a singularity.  The bottom contour containing the horizon ends at $f_0$ and the top
	contour (dotted) is not needed for completeness. (c) The $x$ space representation of the regularized black hole.}
	\label{f2int}
\end{figure}

The singularity in this class of theories  
can be understood using  
 the above horizon/singularity duality.  
 For these black holes, the interior is represented
by a contour in $\tilde f$ space that goes both under and above a branch cut extending to
infinity and ending  at a branch point at some finite $\tilde f=f_0$ (Fig.\,\ref{f2int}(a)). 
The interior solution is self-dual here,
with the above and below the branch contours mapped into each other.  As a result,
the singularity is present (at infinity of the top contour) 
because the horizon is present (at infinity on the bottom contour).

Our strategy to construct a BH solution with a horizon but without a singularity
consists in modifying the nature of the branch point $f_0$ in such a way that
an $\tilde f$ contour going {\em only} under the cut suffices to describe a complete
geometry (Fig.\,\ref{f2int}(b)).  We identify a smaller  class of functions $M(f)$ that can be engineered so that 
the branch point $f_0$ has the desired properties and the singularity is avoided.  The solution is complete, but not self-dual. 
This black hole singularity resolution, which  is 
non-perturbative in $\alpha'$
and closely analogous to the mechanism 
introduced in  \cite{Hohm:2019jgu} to obtain de Sitter vacua in string cosmology, has the following 
general features. 
The interior is again a cosmology with no big-bang singularity at $x=0$. 
In contrast to the standard black hole, however,  it describes an expanding universe that approaches flat 
Minkowski space in 
the infinite future~(Fig.\,\ref{f2int}(c)).  Moreover, the dilaton goes to a fixed
value in that infinite future. 
Such an asymptotic background, however, 
defines a $c=2$ matter conformal field theory (CFT), which is not a 
critical string theory background.\footnote{We thank Ashoke Sen for  
a question that led to this conclusion, and discussions about it.}
It thus appears that 
these regular black holes are solutions of duality-invariant theories,
but apparently not of string theory.

In exploring the geometry of the regular black hole we emphasize that
in a subtle but important way, the $f$ parameterization {\em extends} the definition
of the $\alpha'$ corrections  beyond what the series expansion $f(M)$ can tell us.  Indeed, we are postulating an $M(f)$ that is a well defined function, and for the case of the regular black hole, this function is
{\em not} invertible ($M(f)$ is also not invertible for the cases considered
in~\cite{Gasperini:2023tus}).  Thus, $f(M)$ would have a finite radius of convergence.
The physics of the resolved singularity involves values of $f$ that are not 
achieved within its convergence region.  This allows the late time cosmology 
to have a fixed dilaton instead of a rolling one.  In our re-examination of the derivation
of the equations of motion in the $f$ parameterization (section~\ref{iwltkssvmvg}) we discuss in detail how the issue of non-invertibility affects the analysis.

This paper is organized as follows. In section~\ref{secbh2Dstth} we review the general framework for the 2D black hole and the structure of higher-derivative 
duality-invariant corrections, considering the equations of motion relevant for the exterior and interior geometries.  Section~\ref{thegasvenpar} 
introduces the new Gasperini-Veneziano
parametrization by $f$ and shows how the equations of motion are solved in this framework.
We use the example of the two-derivative BH to understand the basic structure
of branch points and cuts in $f$ space.  The next step is displaying a large class of
solutions that can be viewed as a small general deformation of the two-derivative black hole function $M(f) = -f/2$. We do this in section~\ref{sec_solutions}, where we also examine cubic deformations of the function $M(f)$, which happen not to yield
black holes.  In section~\ref{horandsing} we give the proof that the T-dual of a geometry with a regular horizon must have a curvature singularity.  This result
is used to explain why the back-and-forth contour in $\tilde f$ space used for the interior
solution necessarily leads to an interior with a singularity.  
Finally, in section~\ref{resthebhsing} we find functions $M(f)$ for which the exterior
solution is largely preserved but the interior solution has no singularity. This
is done by adjusting $M(f)$ in such a way that $x$ as a function of $\tilde f$ diverges
at the branch point.  We discuss our results and some possible extensions
in section~\ref{conclremks}.

\section{Black hole in 2D string theory}  \label{secbh2Dstth}

Here we begin by reviewing our notation for the two-dimensional geometry, giving
the general formulae for curvature and the action of T-duality 
on the field variables (for more details see~\cite{Codina:2023fhy}).
We reconsider the two-derivative black hole 
and discuss the equations of motion on the exterior and in the interior, the relevant equations being  
related by a number of sign changes.   Finally, we consider general $\alpha'$ corrections and, again,  
formulate the exterior and interior versions of the 
equations of motion. 

\subsection{Metric, curvature, and duality}\label{sec_generalities}

We consider the target space theory  of bosonic strings in two dimensions (2D)   of Lorentzian signature, 
with the field content restricted to the universal massless sector. Since the 
antisymmetric $b$-field is trivial in two dimensions, we thus include the 
metric $g_{\mu \nu}$ and the  dilaton $\phi$. Denoting coordinates by $x^\mu = (t, x)$, we further 
impose that the fields, and hence the gauge parameters,  do not depend on time $t$. 
This truncated  theory posses a global $O(1,1 ; \mathbb{R})$ 
duality
symmetry together with one-dimensional diffeomorphism invariance: 
reparametrizations of $x$. 
We can assume a diagonal ansatz for the metric
\begin{equation}\label{gmn}
g_{\mu\nu} = \begin{pmatrix} -m^2(x) & 0 \\ 0 & n^2(x) \end{pmatrix}\,,\qquad 
ds^2 = \, - m^2(x) dt^2 \, + n^2(x) dx^2 \,.  
\end{equation}
Here, $m(x)$ and the lapse function $n(x)$ are real. 
Under  coordinate 
transformations of $x$,  the filed $m(x)$ is a scalar while $n(x)$ is a scalar density. The metric $m(x)$ enters the theory through
\begin{equation}\label{M}
M \equiv \frac{D m}{m}\,,
\end{equation}
in terms of the covariant derivative
\begin{equation}\label{D}
D \equiv \frac{1}{n}\frac{d}{d x}\,.
\end{equation}
The scalar dilaton $\phi(x)$ is related to the 
duality invariant dilaton $\Phi(x)$ via
\begin{equation}\label{ord-dil}
\phi(x) = \tfrac{1}{2} \left(\Phi(x) + \log |m(x)|\, \right)\,,    
\end{equation}
which makes the measure of the effective theory to be of the standard form $\sqrt{- g} e^{-2 \phi} = n e^{-\Phi}$. The scalar curvature $R$ of the two dimensional metric
 is given by
\begin{equation}\label{scalarRD}
R = - 2 (M^2 + D M) = - 2\,  \frac{D^2 m}{m}\,.
\end{equation}
The curvature $R$ encodes the full Riemann tensor and hence allows one 
to detect geometric singularities. 

In these 2D backgrounds with a time-like isometry, the duality group acts on the fields as 
\begin{equation}
\label{oddm}
m(x) \to \hat m(x) = \frac{1}{m(x)}\,, \quad n(x) \to \hat n(x) = n(x)\,, \quad \Phi(x) \to \hat \Phi(x) = \Phi(x)\,.
\ee
We will use hatted fields for the fields after duality.  Note that the covariant
derivative $D$ is unchanged by duality.
As a consequence of the transformation of $m$ one also has
\be
\label{MuTD}
M(x) \to \hat M(x) = - M(x)\,,
\ee
The dual scalar curvature, namely, the curvature of the dual metric,  is then given by
\begin{equation}\label{dualRp}
R \to \hat R = -2 ( \hat M^2 +  D\hat M ) = -2 ( M^2 -  DM )\,,
\end{equation}
so we have 
\begin{equation}\label{dualR}
\hat R = -2 ( M^2 -  DM ) = - R - 4 M^2  = R + 4 DM\,.   
\end{equation}

\medskip
\noindent
\textit{\underline{A useful gauge choice}:}
 We can pick a gauge for one-dimensional diffeomorphism invariance by fixing the lapse function $n(x)$. A familiar gauge choice takes  
\begin{equation}
n(x) = 1\,.
\end{equation}
In this gauge the field $M(x)$, if known,  is easily integrated to find $m(x)$: 
\be M(x) = {\partial_x m\over m}  = \partial_x \ln m \quad \Rightarrow  \quad m(x)  =  \exp \Bigl( \int_c^x  M(x') d x' \Bigr)\,, 
\ee
and consequently
\be
\label{mvar}
{m(x_2)\over m(x_1)}=   \exp \Bigl( \int_{x_1}^{x_2}   M(x') d x' \Bigr)\,.
\ee 
The scalar curvature \eqref{scalarRD} now reads
\begin{equation}\label{scalarR}
R = - 2 (M^2 + \partial_xM) = - 2 \,  {\partial^2_x m \over m}\,.
\end{equation}

It is worth emphasizing that all relations above hold for arbitrary 2D metrics of the form (\ref{gmn}), 
independent of any equations of motion, and are hence applicable to generic higher-derivative theories.

\subsection{The two-derivative, two-dimensional black hole} \label{2s2der} 
Before considering $\alpha'$ corrections, 
in the following we recall  the 2D string equations with up to two derivatives and review 
the black hole (BH) solution, both for the exterior region and the interior region. 
We will set up the solutions, both in the $n=1$ gauge in such a way that the 
exterior solution is in the coordinate interval $(-\infty, 0)$ and the interior
solution uses the coordinate interval $(0,\pi)$.  While in both cases we use
the label $x$ to denote the coordinate, these two solutions do not form together
a single solution over the full real line; the two $x$'s are really different.

The target space action for the above string backgrounds reads to leading order 
\begin{equation}\label{I0}
I = \int dx \, n\,  e^{-\Phi} \left[1 +  (D\Phi)^2 - M^2\right]\,, 
\end{equation}
where we use a unit-free notation that absorbs factors of $\alpha'$, denoted by a bar in \cite{Codina:2023fhy} that 
we drop in this paper in order not to clutter the equations. 
Sometimes we refer to the  zero-derivative term as the cosmological term, although it is not a cosmological constant in the standard sense. 
The equations of motion for $m, \Phi$ and $n$ are equivalent to 
\begin{subequations}\label{EOM0}
	\begin{align}
	\frac{d}{dx}\left(e^{-\Phi} M\right) &= 0\,,\\
	\left(\frac{d \Phi}{d x}\right)^2 - M^2 - 1&= 0\,,\\
	\frac{d^2 \Phi}{d x^2} - M^2 &= 0\,, 
	\end{align}
\end{subequations}
where we picked $n = 1$ gauge after performing the variation.

\noindent
\textit{\underline{Exterior BH solution}:}
The above equations admit a unique solution describing  the outside region of a 2D  BH \cite{Witten:1991yr, Mandal:1991tz, Giveon:1991sy}. 
We use the region $x\in (-\infty, 0)$, and 
the metric and duality-invariant dilaton take the form
\begin{equation}\label{BH0}
m(x) = -\tanh \tfrac{x}{2}\,, \qquad \Phi(x) = - \log |\sinh x| + \Phi_0\,,\ \ \ \ 
x< 0 \,, 
\end{equation}
with $\Phi_0$ an integration constant, and the sign of $m(x)$ is fixed to get
the familiar asymptotic behavior.
The scalar curvature \eqref{scalarR} is given by
\begin{equation}
R(x) = \frac{1}{\cosh^2 \frac{x}{2}}\,.
\end{equation}
Here  $x=0$ is the position of the horizon,  $x\rightarrow -\infty$
is the asymptotically flat (AF) region with $m$ approaching one.\footnote{The solution can also be used for $x \in (0, \infty)$, but our later work is simplified by this choice of range of $x$.} Indeed, at $x=0$ the metric vanishes and the curvature is finite
\begin{equation}
m(0) = 0\,, \qquad R(0) = 1\,,
\end{equation}
while in the AF
region  we have
\begin{equation}
\lim\limits_{x \rightarrow -\infty} m(x) = 1\,, \qquad \lim\limits_{x \rightarrow -\infty} R(x) = 0\,.
\end{equation}
On the other hand, the duality-invariant dilaton $\Phi(x)$ becomes infinite
 at both extremes
\begin{equation}
\lim\limits_{x \rightarrow 0^-} \Phi(x) = \infty\,, \qquad \lim\limits_{x \rightarrow -\infty} \Phi(x) = - \infty\,.
\end{equation}
This does not necessarily indicate a pathology since $\Phi$ is not the scalar dilaton $\phi$  that determines the string coupling as $g_s=e^\phi$. Using \eqref{ord-dil}, we get
\be\label{ord-dil_std}
\phi = - \tfrac{1}{2}\log \cosh^2 \tfrac{x}{2} + \phi_0 \quad \Rightarrow \quad  \phi(0) = \phi_0 = \tfrac{1}{2}\left(\Phi_0 - \log 2\right)\,,
\ee
and so the dilaton and hence the string coupling are finite 
at the horizon. 
We also infer $\lim_{x\rightarrow\infty}\phi(x) = -\infty$, so the string coupling $g_s$ goes to zero in the AF region.

\medskip
\noindent
\textit{\underline{Interior BH solution}:}
The solution \eqref{BH0} describes only the exterior region of the BH. Therefore, it has no information on the BH singularity, which lies in the interior. This region, together with others, can be obtained by analytic extension of the exterior solution. Moreover, for the 2D BH, it is possible to obtain the maximally extended solution using a single patch, by the so-called Kruskal-Szekeres coordinates. This coordinate system covers not only the exterior and interior region of the BH, but also a white hole region and a disconnected asymptotically flat space. Since we are mainly interested in the BH singularity, for now it is sufficient to describe the interior region only.

The most straightforward approach to get the interior solution is to look for a transformation that 
preserves the measure and 
changes the metric signature or, more precisely, that maps a metric with time-like isometry to a metric with space-like isometry as follows: 
\begin{align}
n e^{-\Phi}   \  &\to  \ \tilde n e^{-\tilde \Phi}\,,\\
ds^2 = - m^2 (x) dt^2  + n^2(x) dx^2 \quad  &\to \quad ds^2 = \tilde m^2 (x) dt^2  - \tilde n^2(x) dx^2 \,.
\end{align}
This can be achieved by setting 
\be\label{imin}
m =  i \tilde m \,, \quad n = - i \tilde n\,, \quad \Phi = \tilde \Phi - \log i\,,
\ee
which preserves the measure and the product $mn$. Therefore, from \eqref{M} and \eqref{D} we have
\be
D =  i \tilde D \,, \ \ \ M =   i\tilde M \,,
\ee
with  $\tilde D = {1\over \tilde n} {d\over dx}$ and 
$\tilde M = {1\over \tilde m\tilde n} \partial_x \tilde m$. 
Applying these substitutions to the curvature \eqref{scalarRD} we find  $R\to \tilde R$ with
\begin{equation}\label{Rintx}
\tilde R = 2(\tilde M^2 + \tilde D \tilde M) = \, 2 \, \frac{\tilde D^2 \tilde m}{\tilde m}\,.
\end{equation}
Fixing the gauge $\tilde n(x) = 1$,
\begin{equation}\label{Rint}
\tilde R = 2(\tilde M^2 + \partial_x \tilde M) = \, 2 \, \frac{\partial_x^{2} \tilde m}{\tilde m}\,.
\end{equation}
The scalar dilaton in \eqref{ord-dil} is mapped to
\begin{equation}
\tilde{\phi}(x) = \tfrac{1}{2}\left(\tilde \Phi(x) + \log |\tilde m|\right)\,. 
\end{equation}
From the original action \eqref{I0}, we obtain, 
\be\label{Iint}
\tilde I = \int dx \, \tilde n e^{-\tilde \Phi} \left[1 - (\tilde D \tilde \Phi)^2  + \tilde M^2\right]\,.
\ee
Varying this action and fixing the gauge $\tilde n = 1$, the equations of motion read
\begin{subequations}\label{EOM0i}
\begin{align}
\frac{d}{dx}\left(e^{-\tilde \Phi} \tilde M\right) &= 0\,,\\
\Bigl(\frac{d \tilde \Phi}{d x}\Bigr)^2 - \tilde M^2 + 1&= 0\,,\\
\frac{d^2 \tilde \Phi}{d x^2} - \tilde M^2 &= 0\,.
\end{align}
\end{subequations}
The solution in this case is given by trigonometric functions
\begin{equation}\label{BH0i}
\tilde m(x) = \tan \tfrac{x}{2}\,, \quad \tilde \Phi(x) = - \log \sin x + \Phi_0 \quad \Rightarrow \quad 
\tilde{\phi}(x) = -\tfrac{1}{2} \log \cos^2 \tfrac{x}{2} + \phi_0\,,
\end{equation}
where we pick the same integration constant $\Phi_0$ as for the exterior solution in \eqref{BH0}, and hence we got the same $\phi_0$ as in \eqref{ord-dil_std}. The curvature is obtained by inserting  this solution into the right-hand side of \eqref{Rint}: 
\begin{equation}\label{Rintstd}
\tilde R(x) = \frac{1}{\cos^2 \frac{x}{2}}\,.  
\end{equation}
This solution is now valid for the finite range $x \in(0, \pi)$. For $x=0$, the metric, scalar dilaton and curvature take the values
\begin{equation}
\tilde m (0) = 0\,, \qquad \tilde{\phi}(0) = {\phi}_0\,, \qquad \tilde R(0) = 1\,,
\end{equation}
which is consistent with a horizon interpretation. More importantly, the curvature and scalar dilaton on both sides match. At the other end  $x=\pi$, all fields and curvature diverge
\begin{equation}
\tilde m(\pi) = \infty\,, \qquad \tilde{\phi}(\pi) = \infty\,, \qquad \tilde R(\pi) = \infty\,,
\end{equation}
signaling that this is the position of the BH singularity.

\medskip
\noindent
\textit{\underline{Dual BH solutions}:}
As shown in \cite{Giveon:1991sy}, under  T-duality the exterior region of the 2D BH is mapped to a region beyond the singularity, while the interior region is mapped into itself. Here we revisit briefly these phenomena.

Starting from the exterior solution \eqref{BH0} and performing the duality transformation \eqref{oddm}, one obtains 
the new solution, also valid for $x\in (-\infty, 0):$  
\begin{equation}\label{BH0dual}
\hat m(x) = -\coth \tfrac{x}{2}\,, \quad
\hat\Phi(x) = - \log |\sinh x| + \Phi_0\,, \quad \hat \phi (x) = -\tfrac{1}{2} \log \sinh^2 \tfrac{x}{2} + \phi_0\,,
\end{equation}
where the dual metric is the inverse of the original one and the duality-invariant dilaton remains the same. Inserting the dual metric into the dual 
curvature~\eq{Rint}
we get
\be\label{Rxstddual}
\hat R(x) =  - \frac{1}{\sinh^2 \tfrac{x}{2}}\,. 
\ee
In this dual solution, the former horizon $x=0^-$ is mapped to a curvature singularity
\begin{equation}
\lim_{x \rightarrow 0^-} \hat m(x) = \infty\,, \quad \lim_{x \rightarrow 0^-} \hat \phi(x) = \infty\,, \quad \lim_{x \rightarrow 0^-} \hat R(x) = - \infty\,,
\end{equation}
while the asymptotic region is again a flat space-time which lies beyond the BH singularity.

The dual solution to the interior solution \eqref{BH0i} is given by 
\begin{equation}
\hat{\tilde{m}}(x) = \text{cot} \tfrac{x}{2}\,, \quad \hat{\tilde{\Phi}}(x) = - \log \sin x + \Phi_0\,, 
\quad \hat{\tilde{\phi}}   (x) = -\tfrac{1}{2} \log \sin^2 \tfrac{x}{2} + \phi_0\,,
\end{equation}
together with the curvature 
\begin{equation}
\hat{\tilde{R}}(x) = \frac{1}{\sin^2 \frac{x}{2}}\,.
\end{equation}
This geometry still corresponds to the interior region, with the horizon and the singularity exchanged.  This can be made clear by noting that\footnote{We thank M.~Gasperini and G.~Veneziano for pointing this out.} duality plus the `time reversal'
transformation $x \to \pi - x$ leaves the interior solution invariant: $\hat{\tilde{m}}(\pi-x) = \tilde{m}(x)$.

\subsection{General $\alpha'$ corrections} \label{secgenalpha}

In \cite{Codina:2023fhy} we considered  higher-derivative modifications to \eqref{I0} and pointed out that, due to the 
presence of the cosmological term, contributions with more derivatives are generally not subleading relative to 
terms with less derivatives. A naive perturbative expansion is therefore not meaningful unless one assumes 
that the numerical coefficients of the higher-derivative terms obey certain fall-off conditions so that terms with more 
derivatives are  subleading after all. Under this assumption we classified in \cite{Hohm:2019jgu} 
the higher-derivative terms modulo field redefinitions that preserve the fall-off behavior.  
This leads to the minimal 
all-order action 
\begin{equation}\label{Iall}
I = \int dx \, n\,  e^{-\Phi} \biggl[ 1  
+ (D\Phi)^2 + F(M)\biggr]\,,
\end{equation} 
where 
\begin{equation}
F(M) \equiv \sum_{i = 0}^{\infty} \epsilon_i M^{2i + 2} = - M^2 + \epsilon_1 M^4 + \dots\,, \quad \epsilon_0 \equiv -1\,. 
\end{equation}
The fall-off conditions we assumed amount to  the following constraints for the coefficients $\epsilon_i$
\begin{equation}\label{condition}
\epsilon \equiv \epsilon_1 \ll 1\,, \quad \epsilon_i \sim (\epsilon)^i\,,  \  i \geq 1 \,, \quad \epsilon_{p} \epsilon_{k} \sim \epsilon_{p + k}\,, 
\end{equation}
where $\sim$ indicates equality up to numerical constants  of order one.

Taking the variation of \eqref{Iall} with respect to $m, \Phi$ and $n$, combining the dilaton and lapse equations and fixing $n=1$, we arrive at the all-order extension of \eqref{EOM0}
\begin{subequations}\label{EOM}
\begin{align}
\frac{d}{dx}\left(e^{-\Phi} f(M)\right) &= 0\,, \label{EOMM}\\
\left(\frac{d \Phi}{d x}\right)^2 + \check g(M) - 1&= 0\,,
\label{EOMn}\\
\frac{d^2 \Phi}{d x^2} + \tfrac{1}{2} M f(M)&= 0\,. \label{EOMPhi}
\end{align}
\end{subequations}  
Here we introduced\footnote{In our previous work~\cite{Codina:2023fhy},   
we used $g(M)$ for what
is now written as $\check g (M)$.  This change of notation is useful to deal
with the issue of non-invertibility of $M(f)$.} 
\be
\begin{split}\label{fg}
f(M)  &\equiv 
\  F'(M) = \sum_{i=0}^\infty  (2i+2) \epsilon_i M^{2i+1} = - 2 M + 4 \epsilon M^3 + \mathcal{O}(\epsilon^2) \,, \\
\check g(M)
& \equiv \sum_{i=0}^\infty  (2i+1) \epsilon_i M^{2i+2}   = - M^2 + 3 \epsilon M^4 + \mathcal{O}(\epsilon^2)\,,
\end{split}
\ee
which satisfy the following relation
\be\label{g'mf'}
\check g'(M) = M f'(M)\,,
\ee
where ${}^\prime$ denotes the derivative with respect to $M$. 

For the two-derivative theory it was straightforward  to solve \eqref{EOM0} to obtain  \eqref{BH0}, 
the exterior region of the  BH. While it is much harder to find solutions of \eqref{EOM} for generic $f(M)$ and $g(M)$, 
assuming that we have such a solution describing the exterior of a BH, 
we can obtain the solution describing the interior by performing the `signature change' discussed above. 
To this end we apply \eqref{imin} to \eqref{Iall} to get
\be
\tilde I = \int dx \, \tilde n e^{-\tilde \Phi} \left[ 1 - (\tilde D \tilde \Phi)^2  + \tilde F(\tilde M)\right]\,,
\ee
where we have defined
\be 
\tilde F(\tilde M) \equiv   F (i \tilde M) = \sum_{i = 0}^{\infty} (-1)^{i+1} \epsilon_i \tilde M^{2i + 2} = \tilde M^2 + \epsilon \tilde M^4 + \dots
\ee
The equations of motion in the $\tilde n =1$ gauge are now given by 
\begin{subequations}\label{EOMint}
\begin{align}
\frac{d}{dx}\left(e^{-\tilde \Phi} \tilde f(\tilde M)\right) &= 0\,, \label{EOMintM}\\
\Bigl(\frac{d \tilde \Phi}{d x}\Bigr)^2 - \tilde {\check g}(\tilde M) + 1&= 0\,,
\label{EOMintn}\\
\frac{d^2 \tilde \Phi}{d x^2} + \tfrac{1}{2} \tilde M \tilde f(\tilde M)&= 0\,. \label{EOMintPhi}
\end{align}
\end{subequations}
The new function $\tilde f (\tilde M)$ is defined analogously to  \eqref{fg}: 
\begin{equation}
\tilde f (\tilde M) \equiv  \tilde F' (\tilde M)\,,
\end{equation}
while $\tilde {\check g} (\tilde M)$ can be defined via the identity 
\begin{equation}
\tilde {\check g}'(\tilde M) = \tilde M \tilde f'(\tilde M)\,.
\end{equation}
One can check that these functions are related to the ones in  \eqref{fg} via
\begin{equation}\label{tilfeff}
\tilde f (\tilde M)  =  i f (i\tilde M)\,, \quad
 \tilde {\check g} (\tilde M) =  {\check g}(i \tilde M)\,.
\end{equation}
Equations~\eqref{EOM} and \eqref{EOMint} govern the  exterior and interior
of higher-derivative BH solutions, respectively.

For later purposes we need to find out how the inverse functions
$f^{-1}$ and $\tilde f^{-1}$ are related: if $f^{-1}$ is known, what is
$\tilde f^{-1}$?  
We claim that acting on a variable $y$ we have
\be\label{inverse_f}
\tilde f^{-1} (y)  =  -i f^{-1}  (-iy)\,.
\ee 
 This can be checked using the first relation in~\eq{tilfeff} as follows:
\be
\tilde f^{-1} ( \tilde f (\tilde M)) = -i f^{-1} (-i \tilde f (\tilde M))
= -i f^{-1} (-i \cdot i  f (i\tilde M))=   -i f^{-1} (  f (i\tilde M)) = - i \cdot i \tilde M = \tilde M\,. 
\ee 
Noting that we have $f(M)$ and $\tilde f (\tilde M)$ functions, the inverse $f^{-1}$ is the function $M(f)$ and the inverse $\tilde f^{-1}$ is the function $\tilde M (\tilde f)$. With this notation,~\eqref{inverse_f} reads
\be
\label{tildeMM}
\tilde M (\tilde f )  =  -i M (-i\tilde f)\,. 
\ee

\section{The Gasperini-Veneziano parametrization} \label{thegasvenpar}

In order to study solutions of \eqref{EOM} and \eqref{EOMint} 
we will use a useful parameterization inspired by recent work by Gasperini and Veneziano in the context of pre-big bang string cosmology~\cite{Gasperini:2023tus}.  Instead of describing the fields as functions of the
coordinate $x$, they are described as functions of a new parameter, which arises
as follows.  
One considers the function $f(M)$ in \eqref{fg}, a 
typically  
infinite series in the metric variable $M$,  and inverts it to find $M(f)$.  
In $M(f)$
the metric variable is parameterized by~$f$, the Gasperini-Veneziano parameter.  
One can then find $x(f)$ by performing an integral analytically, or numerically if needed. 
We will distinguish the parameter for each region, using $f$ to describe the exterior and $\tilde f$ for the interior.

There is a shift in perspective here. While one originally would view the theory
space as described by the set of possible functions $f(M)$, we now consider
the theory space as described by the set of functions $M(f)$. 
The invertibility issues are actually central to the development
and we will argue that the $f$ parameterization leads to a natural
{\em extension} of the original equations of motion, beyond their original
domain of validity.  Clarifying this will
be our first task in this section.

In this section we show how to solve the exterior and interior equations of 
motion using this parameterization.  We then do this explicitly for the 
two-derivative BH.  Here the nature of the contours required in $f$ space becomes
evident. 

\subsection{Multivalued functions and equations of motion} \label{iwltkssvmvg} 

As mentioned above and reviewed in section~\ref{secgenalpha}, the conventional definition of $\alpha'$ corrections begins by stating that the 
Lagrangian contains a function $F(M)$ given as an infinite power series expansion 
in (even) powers of $M$, each successive term containing an additional
power of $\alpha'$.  It is quite possible, perhaps even likely, that this series
only has a finite radius of convergence, making the definition of the action
incomplete.  From $F(M)$ one defines $f(M) \equiv F'(M)$ 
and one usually works with $f(M)$.  The viewpoint adopted here is that
$\alpha'$ corrections are described by the {\em function} $M(f)$ which we assume
to be well defined (single valued), and that
$f(M)$ is just its inverse.  
Thus we have 
\be
M(f)\ \hbox{ is a well defined function (single valued).}
\ee
It follows that the inverse $f(M)$ is usually multivalued.
Indeed, assume $M(f)$ grows monotonically 
 from zero at $f=0$ to some maximum $M_*$ at some
value $f_*>0$,  and then falls down.  Then the inverse $f(M)$ is not single valued.
In fact if one expands $M(f)$ as a series in $f$ and perturbatively inverts it,
 $f(M)$ will only converge up to $M= M_*$, where it has a branch point.  
 In general, let $M_*$ denote the 
 maximum value of $M$ 
 for which the series $f(M)$ converges. 
  In summary,  we have two important additional facts: 
  \begin{quote}
  The inverse $f(M)$ of $M(f)$ is not single valued. The series definition of $f(M)$ converges for $M \in [0, M_*]$ and $f(0)= 0$.
  \end{quote}
The convergent series is a precise definition of $f(M)$ in this range. 

The equations of motion also feature a quantity $\check g (M)$ defined by two relations 
\be
{d\check g \over dM} =  M {df\over dM}\   \ \ \hbox{and} \ \ \ \check g (0) = 0\,.
\ee
Both of these are implemented by the integral definition
\be
\check g (M) \equiv \int_{u=0}^{u=M}  u  {df(u)\over du} du\,, \ \ \  M \in [0, M_*]\,.
\ee
The condition on the range of validity of the definition is needed because
 $\check g$ is a multivalued function.  Indeed,  $df/du$ appearing inside
the integral, is multivalued, a property it inherits from 
$f(u)$.  But now, in this range we can change the variable of integration
from $u$ to $f$.  In doing so $u$ must be thought as $u(f)$, where this
is the inverse function of $f(u)$.  
The integral then becomes
\be
\check g (M) \equiv \int_{f(u=0)}^{f(u=M)}  u(f) df =  
\int_{0}^{f(M)}  u(f') df'  \,, 
\ee
so that, 
recalling that we call $M$ the inverse function of $f$, 
\be
\check g (M) \equiv  \int_{0}^{f(M)}  M(f') df' 
\,, \ \ \  M \in [0, M_*]\,.
\ee
The above definition suggests the following
construction of a $g(f)$ well defined for {\em all} $f$:
\be
g(f) \equiv  \int_0^f  M(f') df'  \,.
\ee
It follows immediately from the last two equations that  
\be
\label{extentsiong}
\check g (M) = g(f (M))   \, , \  \ \ \, M \in [0, M_*]\,.
\ee
This states that in the domain of definition, $\hat g (M)$ is
given by the well-defined $g(f)$. 

Let us now consider the field equations, following the steps considered in~\cite{Gasperini:2023tus}, but 
in the light of the above observations.  
In the two-dimensional theory we have~\eq{EOMn} that reads
\be
\Bigl( {d\Phi\over dx} \Bigr)^2  = 1 -\check g (M) \,,  \ \ M \in [0, M_*] \,.
\ee
Using the above definition of $\check g$ we have
\be
\Bigl( {d\Phi\over dx} \Bigr)^2  = 1 -  g (f(M)) \,, \ \ M \in [0, M_*] \,.
\ee
Here the dilaton derivative is effectively set equal to a function of $f(M)$.
But then we may as well forget $M$,  declaring that this equation
 sets the dilaton derivative equal to a {\em function of~$f$}:
\be
\label{eorid38}
\Bigl( {d\Phi\over dx}\Bigr)^2  =  1 - g (f)  =  1 - \int_0^{f} M(\tilde f)   d \tilde f\, , \ \  \ \ \  \forall \, f \,.    
\ee
This equation is now valid for {\em all} $f$, thus going
beyond the limitations of the power series
definition of the $\alpha'$ corrections.  
The equation indeed yields a parameterization of the dilaton derivative in terms of $f$:
\be\label{dPhiout}
{d\Phi\over dx } (f) = \pm \sqrt{P(f) }\,,  \ \ \hbox{with} \ \ P(f) \equiv  1 -{\int_0^{f} }  M(f') d f' \,. 
\ee
The plus/minus signs correspond to using different branches of the square root.
The equations of motion 
also imply that  $f$ ends up being a coordinate, which can be
related to the $x$~coordinate. Indeed, consider~\eqref{EOMM} and 
distribute the $x$ derivative 
\be\label{aux1}
{d\Phi\over dx} f(M) =  {df\over dx}  \quad \Rightarrow 
\quad  dx = {df\over f} {1 \over {d\Phi\over dx} (f)} \,.
\ee
Using \eqref{dPhiout},  
the second equality leads to an equation for the coordinate $x$ as a function of $f$
\be
\label{x-diffe}
dx = \pm  {df\over f  \sqrt{P(f) }} \,. 
\ee
Finally, consider again~\eqref{EOMM}, which states that the combination
$e^{-\Phi} f(M)$ is
a constant relative to the $x$ coordinate.  It is therefore a constant relative
to $f$, and therefore,  
\begin{equation}\label{dil-f}
\Phi(f) = \log |f| + \Phi_1\,,
\end{equation}
with $\Phi_1 \equiv \Phi(f=1)$ an integration constant that absorbs the sign of $f$ and encodes the mass of the black hole.

The above equations 
represent an extension of the original equations,
and we will assume these provide the non-perturbative 
definition  of the theory with $\alpha'$ corrections.

\subsection{Parametrization in terms of $f$}\label{sec_f}  
In this section  we consider the equations of motion in the $f$
parameterization, both for the exterior and for the interior.
Note that we have passed from using $\check g(M)$ to using $g(f)$, in the exterior,
and from using $\tilde{\check g}(\tilde M)$ to using $\tilde g(\tilde f)$ in the interior.
The analog of~\eq{eorid38}  for the interior follows from~\eq{EOMintPhi} and becomes
\be
\label{eorid39}
\Bigl( {d\tilde\Phi\over dx}\Bigr)^2  =  -1 + \tilde g (\tilde f)  \,. 
\ee

\noindent
\textit{\underline{Exterior solution}:}  Much of the work was already
done above, so we just make some comments, consider a few extra equations,
and collect results. 
Consider again~\eq{dPhiout}:
\be 
{d\Phi\over dx } (f) = \pm \sqrt{P(f) }\,,  \ \ \hbox{with} \ \ P(f) \equiv  1 -{\int_0^{f} }  M(f') d f' \,. 
\ee
The allowed range of $f$ is determined by the condition $P(f) \geq 0$.
We see that $P(f = 0) = 1 > 0$, so by continuity there is at least an interval around $f=0$ where the solution exists.  Moreover, if the integral appearing above is negative for all values of $f$ (as 
will be the case for the standard BH), then the solution is valid for the whole real line $f \in \mathbb{R}$.

Using \eqref{x-diffe} in \eqref{mvar} we obtain an expression for the metric components
\be
\label{mvarx}
{m(x_2)\over m(x_1)}=  {m(f_2)\over m(f_1)} =    \exp \biggl( \pm\int_{f_1}^{f_2}  {df\over f}  {M(f) \over\sqrt{P(f)} } ~\biggr) \,.  
\ee 
With $m(f)$, we can relate $\Phi(f)$ to the scalar dilaton $\phi(f)$ via \eqref{ord-dil}, where now all quantities are parameterized by $f$. 
For the curvature, we have from~(\ref{scalarR}), 
\be
\label{Rout}
R_\pm (f) = -2 \Bigl(  M^2 + {d M\over df} {d f\over  dx} \Bigr)  
=  -2 \Bigl(  M^2 \pm   f \, {d M\over d f}  
\sqrt{ P(f) } \ \ \Bigr)  \,. 
\ee
Collecting the results, for ease of reference, we have
\begin{subequations}\label{fexte}
\begin{align}
{d \Phi\over dx } ( f) &= \ \pm \sqrt{ P(f) }\,, \quad \quad 
\Phi( f) = \log | f| + \Phi_1\,, \label{fextP}\\
dx &= \ \pm  {d f\over  f  
\sqrt{  P( f) }}\,, 
\label{fextxf}\\
{m(x_2)\over m(x_1)} &=  {m(f_2)\over m(f_1)} =    \exp \biggl( \pm\int_{f_1}^{f_2}  {df\over f}  {M(f) \over\sqrt{P(f)} } ~\biggr) 
\,, \label{fextm}\\
R_\pm (f) &= -2 \Bigl(  M^2 + {d M\over df} {d f\over  dx} \Bigr)  
=  -2 \Bigl(  M^2 \pm   f \, {d M\over d f}  
\sqrt{ P(f) } \ \ \Bigr)  \,.  \label{curvxext}
\end{align}
\end{subequations}
Here the argument $P ( f)$ of the square roots is 
\begin{equation}\label{Pfext}
 P(f) \equiv 1 - \int_0^{f}  M( f') d f'\,.
\end{equation}
Equations~\eqref{fexte} 
encode the general solution to \eqref{EOM} in terms of $f$ as a parameter. 
The solution is completely determined once an ansatz for $M(f)$ is given, and the physics depends on such choice.  To make this point more emphatically, we can write the metric $ds^2 = -m^2(x) dt^2 + dx^2$ by passing from the $x$ coordinate to the $f$ coordinate using~\eq{fextxf},
and writing $m(x)$ using~\eq{fextm}:  
\be
\label{metricinfspace}
ds^2 = -  \exp\Biggl(  \pm\,  2 \int^f {du\over u} {M(u)\over \sqrt{P(u)}}  \Biggr)  \, (dt)^2 
\ + \ {(df)^2 \over f^2 P(f) } \,, \ \ \ \Phi(f) = \log|f| + \Phi_1 \,,
\ee
where we also included the dilaton.  It may be that eventually $x$ can be traded
for $f$, but we will not explore this here.  Still, the above makes clear that we
have a solution if we know $M(f)$.

Without $M(f)$ we cannot say much about the specifics of each solution, but we can still analyze some global aspects of them: the first thing to notice is that, from our starting point in \eqref{fg}, 
since $f(M)$ has a power series expansion in $M$, its inverse $M(f)$, at least
perturbatively,  is expected to have the following expansion
\begin{equation}\label{Mf_expansion}
M(f) = -\tfrac{1}{2} f\left[1 + \frac{\epsilon}{2} f^2 + \mathcal{O}(\epsilon^2 f^4)\right] = -\tfrac{1}{2}\,  f\left[1 + h(f^2)\right]\,,
\end{equation}
where we introduced the
function $h$ as follows
\begin{equation}\label{hxi}
h(\xi) \equiv \frac{\epsilon}{2} \,\xi + \mathcal{O}(\epsilon^2 \xi^2)\,.
\end{equation}
Here $M(f) = - \frac{f}{2}$, obtained for  $h(\xi) = 0$,  corresponds to the standard two-derivative case, for which we recover the exterior BH solution of \eqref{BH0}, as we will show in 
section~\ref{sec_standardBH}. 
The expansion \eqref{Mf_expansion} implies the following two properties for generic 
$M(f)$
\begin{equation}\label{Mf0}
M(f=0) = 0\,, \quad M(-f) = - M(f)\,,
\end{equation}
and 
\begin{equation}
h(\xi=0) = 0\,, \quad h'(\xi=0) = \text{finite}\,.
\end{equation}
These observations have 
the important consequence that $f=0$ corresponds to the faraway region in all solutions. 
To see this we study the behavior of~\eq{fexte} and~\eq{Pfext}
near 
$f=0$ 
for generic $M(f)$. Using \eqref{Mf_expansion}, \eqref{Mf0} and $P(f=0)= 1$, we get
\begin{equation}
\Phi(0) = - \infty\,, \quad dx \simeq \pm \frac{df}{f}\,, \quad  m(f) \simeq m(0) e^{\mp \frac12 f}\,, \quad R(0) = 0\,.
\end{equation}
From the second relation we infer
$x\simeq \pm \ln f+{\rm const}$ 
and hence  
that $f=0$ corresponds to an asymptotic 
region with infinite 
$x$. 
From the third relation we see that we can choose the integration constant such that $m(0) = 1$. Using this together with $\Phi(0) = - \infty$ in \eqref{ord-dil} we obtain $\phi(0) = - \infty$. All these results are compatible with the 
interpretation as a faraway region.

This makes $f=0$ the end of the space-time, which implies that solutions 
with $f < 0$ and with $f > 0$ should be treated separately, not as two regions of the same exterior solution. On top of this distinction, we have the two branches of the square root of \eqref{dPhiout} corresponding to the $\pm$ choices. While this seems to suggest we have four different solutions, all of them 
should be physically equivalent in string theory. 
More precisely, the negative and positive regions of $f$ are related by T-duality 
since $M \to - M$ 
and~\eqref{Mf_expansion} imply 
\begin{equation}\label{f-f}
f \to \hat f = - f\,.
\end{equation}
Moreover, solutions with different signs of the square root~\eqref{dPhiout} are connected via a trivial sign flip of the $x$ coordinate. More precisely, the solution with minus sign and $f < 0$ is identical to the solution with plus sign and $f > 0$, upon changing $x \to - x$. Their T-dual solutions can be obtained by~\eqref{f-f}. 

\medskip
\noindent
\textit{\underline{Interior solution}:}

In order to get the interior solutions in the Gasperini-Veneziano  
 parameterization, we repeat the procedure we just performed for the exterior case, but this time using the equations     
         \eqref{EOMint}. 
Without going into details, the solutions  read
\begin{subequations}\label{fint}
\begin{align}
{d\tilde \Phi\over dx } (\tilde f) &= \ \pm \sqrt{\tilde P(\tilde f) }\,, \quad \quad \tilde \Phi(\tilde f) = \log |\tilde f| + \Phi_1\,, \label{fintP}\\
dx &= \ \pm  {d\tilde f\over \tilde f  
\sqrt{ \tilde P(\tilde f) }}\,, 
\label{fintxf}\\
{\tilde m(x_2)\over \tilde m(x_1)} &= {\tilde m_{\pm}(\tilde f_2)\over \tilde m_{\pm}(\tilde f_1)} = \exp \biggl( \pm\int_{\tilde f_1}^{\tilde f_2}  {d f\over f}  {\tilde M(f) \over
\sqrt{\tilde P( f)
} } ~\biggr) \,, \label{fintm}\\
\tilde R_{\pm}(\tilde f) &= 2 \Bigl( \tilde M^2 +  {d\tilde M\over d\tilde f} {d\tilde f\over dx} \Bigr)  
=  2 \Bigl( \tilde M^2 \pm \tilde f \, {d\tilde M\over d\tilde f}  
\sqrt{ \tilde P(\tilde f)  } \ \ \Bigr)\,. \label{curvxint}
\end{align}
\end{subequations}
Here the argument of the square roots, called $\tilde P (\tilde f)$, 
is defined as follows
\begin{equation}\label{Pf1}
\tilde P(\tilde f) \equiv -1 + {\int_0^{{\tilde f}} } \tilde M(\tilde f') d\tilde f'\,.
\end{equation}
Since $\tilde P(\tilde f = 0) = -1 < 0$, this time the solution excludes the point $\tilde f=0$. Moreover, in general it is not even guaranteed that the interior solution exist at all. There would be no interior solution if
$\tilde P(\tilde f)$ is negative for all $\tilde f$.

Using \eqref{tildeMM} we can determine  the 
function
 $\tilde M(\tilde f)$ for the interior region from the 
 function
 $M(f)$ for the exterior region given in \eqref{Mf_expansion}: 
\begin{equation}
\tilde M(\tilde f) = \tfrac{1}{2} \tilde f \left[1 - \tfrac{\epsilon}{2} \, \tilde f^2  + \mathcal{O}(\epsilon^2 \tilde f^4)\right] = \tfrac{1}{2} \tilde f \left[1 + h(-\tilde f^2)\right]\,,
\end{equation}
with the same $h(\xi)$ defined in \eqref{hxi}.

The range of validity of the interior solution is very different from the one of the exterior solution.  From now on we assume there is always at least one positive branch point  
$f_0 > 0$ such that $\tilde P(f_0) = 0$ and $\tilde P(\tilde f) > 0$ for some interval $\tilde f \in (f_0, f_1)$, where $f_1 > f_0$ can be another branch point or
the point at infinity. If this is the case, due to the 
 even
 parity of $\tilde P(\tilde f)$, the interior solution exists at least for an interval $\tilde f \in (-f_1, - f_0) \cup (f_0, f_1)$. However, since for the interior solution T-duality also acts as
\begin{equation}\label{tildef-tildef}
\tilde f \to \hat{\tilde f} = - \tilde f\,,
\end{equation}
both negative and positive regions are related via a duality transformation. 
We are imposing to have at least one branch point for positive $f$. With multiple
such branch points it is possible to have multiple separated domains within $\tilde f > 0$ such that $\tilde P(\tilde f)$ is positive in each of those disconnected regions.

\subsection{The standard black hole}\label{sec_standardBH} 

Here we work out the presentation of the standard two-derivative BH solution of section \ref{sec_generalities} in the 
Gasperini-Veneziano parametrization.  The experience gained here will prove
essential for the later generalizations.

\medskip
\noindent
\textit{\underline{Exterior solution}:} The $M(f)$ ansatz for the exterior region of the two-derivative theory corresponds to the limit  $\epsilon \to 0$ in \eqref{Mf_expansion}:
\begin{equation}\label{Mstd}
M(f) = -\tfrac{1}{2} f\,.
\end{equation}
Associated to $M$ we have $P(f)$ given in~\eqref{Pfext}:
\begin{equation}\label{PPfstd}
P(f) = 1 + \tfrac{1}{4}\,  f^2 \ > \ 0 \ \quad  \forall f \in \mathbb{R}\,,
\end{equation}
which implies that all $f$ are in principle allowed. 
The duality-invariant dilaton does not depend on the specific ansatz for $M(f)$, so it is still given by \eqref{dil-f}. 
Using \eqref{PPfstd} in \eqref{dPhiout}, its $x$-derivative is given by 
\be
\label{dPhix}
{d \Phi\over dx} =  \sqrt{1 + \tfrac{1}{4} f^2} \,,
\ee
where we choose the plus sign to fit conventions 
where $x=0^-$ is the horizon and the exterior region
is $x <0$, with the
AF region at $x \to -\infty$. Moreover, to stick to this convention, we need to choose $f \geq 0$. The relation between $x$ and $f$ can be obtained from \eqref{x-diffe}, which for this ansatz takes the form
\be\label{aux}
dx =  {df\over f} {1\over \sqrt{1 + \tfrac{1}{4} f^2}} = \  - d \, \hbox{ArcSinh} \,
\tfrac{2}{f} \,.
\ee
Integrating,  
\be\label{xfstd}
x(f) = \ -  \hbox{ArcSinh} \tfrac{2}{f} \ \  \to \  \  \tfrac{f}{2}  = -\, \tfrac{1}{\sinh \, x} \,, 
\ee
where we choose the integration constant to be zero. The $x=0^-$ horizon corresponds to $f= \infty$ and the AF region
$x =  -\infty$ corresponds to $f=0^+$:
\be\label{horizonfaraway}
 \hbox{AF region:}  \ x= -\infty\, , \ \   f = 0^+\,,
  \quad  \ \hbox{Horizon:}  \ x= 0,  \ \ f= +\infty \,.
\ee
Increasing $x$ corresponds to increasing $f$.

\begin{figure}[ht]
	\centering
\epsfysize=6.1cm
\epsfbox{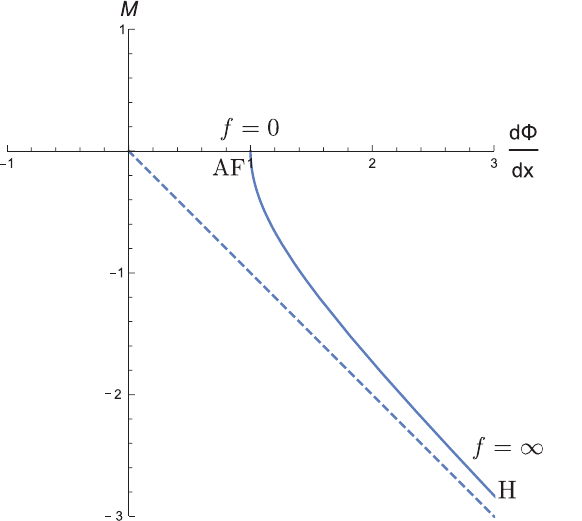}
	\caption{\small Black hole exterior. The solid line is the plot of the dilaton derivative  $d\Phi/dx$ and $M$ as a function of $f \in [0, \infty)$. The point on the horizontal axis is 
	$f=0$ and represents the AF region. As $f\to \infty$ we reach the horizon
	and the curve asymptotes to the dashed line $|M| = |{d\Phi\over dx} |$. }
	\label{figBHext_MP_plots}
\end{figure}

In order to determine 
the metric we take the plus sign in~\eqref{fextm} 
and use~\eqref{Mstd} and~\eqref{PPfstd}: 
\be
\label{mvarxx}
{m(f_2)\over m(f_1)} =    \exp \Biggl( -\frac{1}{2}  \int_{f_1}^{f_2}  {df \over\sqrt{ 1 + {f^2\over 4}} } \Biggr) \,. 
\ee 
We can  integrate this analytically, but before doing so  
it is instructive to consider the ratio of the scale factor at 
the horizon $f=\infty$ divided by the value at the AF region about
$f=0$:
\be
\label{mvarxxx}
{m(f=\infty)\over m(f=0)} =    \exp \Biggl( -\frac{1}{2}  \int_{0}^{\infty}  {df \over\sqrt{ 1 + {f^2\over 4}} } \Biggr)  =  0 \,, 
\ee 
due to the logarithmic divergence at the top limit of integration, and confirming
the vanishing of the metric at the horizon: $m(f=\infty)=0$. 
 Since the scale of $m$ is not detected by the equations of motion, we can set $m(f=0)= 1$.  Setting $f_2=f$ and $f_1=0$  in~\eqref{mvarxx} we get
\be
\label{mfstd}
m(f) = \exp \Biggl( -\frac{1}{2} \int_0^f {df' \over\sqrt{ 1 + { f'^2\over 4}} } \Biggr)  =  \exp \Bigl( -\hbox{ArcSinh} \tfrac{f}{2} \Bigr)
=  -\tfrac{f}{2} + \sqrt{1 + \tfrac{f^2}{4}}\,.
\ee
This indeed gives $m(0)=1$, and $m(f) \simeq 1/f$ for large $f$, so that $m(\infty) = 0$. 
We now compute the curvature from \eqref{Rout}, using the top sign and finding
\begin{equation}\label{Routstd}
\begin{aligned}
R(f)  = \   f\left(  -{f\over 2}  +
\sqrt{ 1 + {f^2\over 4}   }    \right) =  f m(f)\,.
\end{aligned}
\end{equation}
Note that $R(0)= 0$, as befits the AF region, and $R(f) \simeq 1$ for $f\to \infty$, which is the value of the curvature at the horizon. 

At this point we have $M(f), \Phi(f), m(f)$ and $R(f)$, and so the full
$f$-parameterized solution is determined. As a consistency check one may use the relation between $f$ and $x$ to 
recover the BH solution in the familiar form reviewed in section~\ref{2s2der}. 
For the metric, for example, we have, using~\eq{xfstd}, and recalling that $x$ is negative, 
\be
m(x) = \tfrac{1}{\sinh x} + \sqrt{ 1 + \tfrac{1}{\sinh^2 x}} = \tfrac{1}{\sinh x}\bigl( 1 - \sqrt{1 + \sinh^2 x} \bigr) =  - \tanh \tfrac{x}{2} \,. 
\ee
For the dilaton we have, since $f>0$, 
\be
\Phi = \log f + c  =  \log \bigl( \tfrac{ -2 }{\sinh x} \bigr) + c = -\log |\sinh x|+ c'\,. 
\ee

\medskip
\noindent
\textit{\underline{Interior solution}:}

The BH interior solution of \eqref{BH0i} can also be recovered in the $\tilde f$ parameterization, 
using \eqref{fint}. 
The ansatz for $\tilde M(\tilde f)$ is obtained by using the formula \eqref{tildeMM} on the ansatz for the exterior \eqref{Mstd}:
\begin{equation}
\tilde M(\tilde f) = \tfrac12 \tilde f\,.
\end{equation}
The argument of the square root this time takes the form
\begin{equation}
\tilde P(\tilde f) = -1 + \tfrac14 \tilde f^2\,,
\end{equation}
and so the square root has branch cuts on the real line with $|\tilde f|> 2$, with branch points at $\tilde f= \pm 2$. Therefore, the interior solution is valid for the interval $\tilde f \in (-\infty, -2) \cup (2, \infty)$, where negative and positive regions are related via T-duality \eqref{tildef-tildef}. From now on we choose $\tilde f \geq 2$.

In order to describe all of the internal region in the $\tilde f$-parameterization we need to cover the range $\tilde f \geq 2$ twice. We begin from $\tilde f = \infty$ to $\tilde f = 2$ traveling under the branch cut, and then we return from $\tilde f = 2$ to $\tilde f = \infty$ over the cut. The difference between paths is given by the choice of $\pm$ sign in front of the square root in the solution \eqref{fint}. The square root is assumed to take negative values below the cut ($-$ sign) and positive values above the cut ($+$ sign).

When going under the cut, the dilaton derivative is given by
\begin{equation}\label{Pintstd}
\frac{d \tilde\Phi}{d x}(\tilde f) = - \sqrt{-1 + \tfrac14 \tilde f^2} \,.
\end{equation}
The relation between $x$ and $\tilde f$ in \eqref{fintxf} takes the form
\begin{equation}\label{dxdfintstd}
d x = - \frac{d \tilde f}{\tilde f \sqrt{-1 + \tfrac14 \tilde f^2}} = - \frac{d \tilde f}{|\tilde f| \sqrt{-1 + \tfrac14 \tilde f^2}} = d\, \hbox{arccsc}\, \tfrac{\tilde f}{2}\,, \quad \tilde f \geq 2\,. 
\end{equation}
Integrating this equation and setting the integration constant to zero we get
\begin{equation}\label{xfdown}
x(\tilde f) = \hbox{arccsc}\, \tfrac{\tilde f}{2} \quad \to \quad \tfrac{\tilde f(x)}{2} = \csc x = \frac{1}{\sin x}\,.
\end{equation}
In this parameterization
as $\tilde f$ decreases 
$x$ increases. Indeed, we have $x (\tilde f=\infty) = 0$ and $x (\tilde f=2) = \tfrac{\pi}{2}$:
\be
\tilde f\in [\infty, 2]_- \ \to \  x \in [0, \tfrac{\pi}{2}] \,,
\ee
with the minus subscript indicating that we are traveling under the cut. As we will see, $\tilde f=\infty$ ($x=0$) corresponds to the position of the horizon, while $\tilde f = 2$ ($x = \tfrac{\pi}{2}$) is just an intermediate point in the interior solution.

For the metric we need \eqref{fintm} with the minus sign. Choosing the lower boundary to be $\tilde f_1 = 2$ and leaving the upper boundary arbitrary we get
\begin{equation}
\tilde m_{-}(\tilde f) = \tilde m_{-}(2) \exp\Biggl(- \tfrac12 \int_{2}^{\tilde f} \frac{df}{ \sqrt{-1 + \tfrac14 f^2}}\Biggr) = \tilde m_{-}(2) \exp\left(- \hbox{arccosh} \tfrac{\tilde f}{2}\right)\,,
\end{equation}
and since $\hbox{arccosh}(1) = 0$ we can take $\tilde m_{-}(2) = 1$. Since
$\hbox{arccosh}\, y = \log \left(y + \sqrt{-1 + y^2}\right)$, 
we obtain 
\begin{equation}\label{mtildestd}
\tilde m_{-}(\tilde f) = \tfrac{\tilde f}{2} - \sqrt{-1 + \tfrac{\tilde f^2}{4}} = 
\tfrac{\tilde f}{2} \left(1 - \sqrt{- \tfrac{4}{\tilde f^2} + 1}\right)\,,
\end{equation}
where for the second equality we used that $\tilde f$ is positive. For the curvature  
one can show with \eqref{curvxint} that 
\begin{equation}
\tilde R_{-}(\tilde f) = \tilde f \left(\tfrac{\tilde f}{2} - \sqrt{-1 + \tfrac{\tilde f^2}{4}}\right) = \tilde f \tilde m_{-}(\tilde f)\,.
\end{equation}

As anticipated, $\tilde m_{-}(\tilde f)$ and
$\tilde R_{-} (\tilde f)$ are consistent with $\tilde f = \infty$ being the location of the horizon since
\begin{equation}
\lim_{\tilde f \to \infty} \tilde m_{-}(\tilde f) = 0\,, \quad \lim_{\tilde f \to \infty} \tilde R_{-}(\tilde f) = 1\,, 
\end{equation}
as  can be checked with  the expansion around $\tilde f \to \infty$ of the second equality in \eqref{mtildestd}. On the other hand, $\tilde f = 2$ is just an intermediate point in the interior solution such that
\begin{equation}\label{mR2}
\tilde m_{-}(2) = 1\,, \quad \tilde R_{-}(2) = 2\,. 
\end{equation}

The other half of the solution is recovered by going over the cut. In this situation, the dilaton derivative \eqref{fintP} and the relation between $x$ and $\tilde f$ from \eqref{fintxf} have the opposite sign as in \eqref{Pintstd} and \eqref{dxdfintstd}. For the latter we have
\begin{equation}
dx = - d\, \hbox{arccsc}\, \tfrac{\tilde f}{2} = d\, \hbox{arccos}\, \tfrac{2}{\tilde f} \quad \Rightarrow \quad x(\tilde f) = \hbox{arccos}\, \tfrac{2}{\tilde f} + c_0\,,
\end{equation}
where after integration we picked a non-trivial integration constant $c_0$. This is necessary since at $\tilde f=2$ we already had $x = \tfrac{\pi}{2}$ from under the cut, so by continuity we must have
\be
\label{xoveepurcut}
\quad  \tfrac{\pi}{2}  = \hbox{arccos} \, 1 + c_0 = c_0  \quad \to \quad c_0 = \tfrac{\pi}{2} \,.
\ee
Therefore, 
\be
\label{xftop}
x(\tilde f) - \tfrac{\pi}{2}  = \hbox{arccos} \, \tfrac{2}{\tilde f}  \quad \to \quad \tfrac{2}{\tilde f(x)} = \cos (x - \tfrac{\pi}{2} ) = \sin x \,. 
\ee
Indeed we have $x (\tilde f=2) = \tfrac{\pi}{2}$ and $x (\tilde f=\infty) = \pi$, since $x$ grows with $\tilde f$.  So we have
\be
\tilde f\in [2, \infty]_{\tiny +}  \ \to \  x \in [\tfrac{\pi}{2}, \pi] \,,
\ee
where the subscript $+$ indicates going on top of the cut. It is interesting to note that the relation $\tilde f(x)$ is the same for under 
the cut, c.f.~\eqref{xfdown}, and over the cut, c.f.~\eqref{xftop}. The former covers $x \in (0, \tfrac{\pi}{
2})$ and the latter the remaining half of the interior solution $x \in (\tfrac{\pi}{
2}, \pi)$.

\begin{figure}[ht]
	\centering
 \epsfysize=6.0cm
\epsfbox{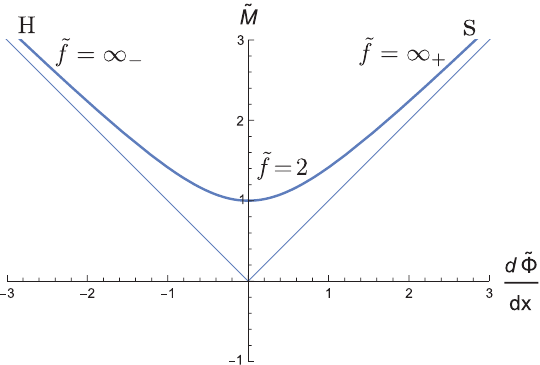}  
\caption{\small Black hole interior. 
The solid line plots the 
relation between the dilaton spatial derivative and $\tilde M$.
The horizon H is on the upper left, the singularity S on the upper right (both at 
$\tilde f=\infty$,
below and above the cut).  
The branch point at $\tilde f=2$ 
corresponds to the point with minimal $\tilde M=1$.  The $\tilde f$ contour is that shown in Fig.\ref{f1int} (b).  }   
\label{bh-interior-nfig} 
\end{figure}

This time, taking the plus sign choice in \eqref{fintm} and \eqref{curvxint}, we obtain the expressions for the metric and curvature over the cut:
\begin{equation}
\tilde m_+(\tilde f) = \tfrac{\tilde f}{2} + \sqrt{-1 + \tfrac{\tilde f^2}{4}}\,, \quad \tilde R_+(\tilde f) = \tilde f \tilde m_+(\tilde f)\,. 
\end{equation}
The intermediate point $\tilde f = 2$ ($x = \tfrac{\pi}{2}$) gives again~\eqref{mR2}, but this time the boundary $\tilde f = \infty$ ($x=\pi$) corresponds to the singularity, where
\begin{equation}
\lim_{\tilde f \to \infty} \tilde m_{+}(\tilde f) = \infty\,, \quad \lim_{\tilde f \to \infty} \tilde R_{+}(\tilde f) = \infty\,. 
\end{equation}
As a consistency check one may again verify that the solution in the form \eqref{BH0i} follows from 
the relation between $\tilde f$ and $x$.  A plot of the dilaton derivative $d\tilde \Phi/dx$  and the metric derivative $\tilde M$,  both as a function of $\tilde f$ helps
visualize the solution~(Fig.~\ref{bh-interior-nfig}).  As we go from the horizon
to the singularity the dilaton derivative first decreases and then increases.  The metric
$\tilde m$ goes from zero to infinity, while $\tilde M$ first decreases and then increases.

\section{$\alpha'$ corrected black hole solutions}\label{sec_solutions}  

In the first part of this section   
we present a class of functions $M(f)$ that lead to BH solutions in the
conventional sense.  These solutions have an exterior and an interior region. The exterior includes in $f$-space 
the point zero, 
which can be identified with the AF region,
and the point at infinity, 
which can be identified with a horizon. 
The interior region begins at a horizon and ends in a singularity, that is not removed in this class of solutions. Both regions, we believe,  form a single solution because the curvature on both regions attain the same value at the horizon.  We make no claims
that this is the most general class of $M$'s that lead to conventional BH solutions.

We then consider a cubic ansatz for $M(f)$, the simplest deformation of the 
expression $M(f) = -\tfrac{1}{2}f$ valid in the two-derivative theory. 
We do not find black hole
solutions here. 
Instead we find situations where the exterior and interior solutions
do not seem to form a single solution, situations 
with naked singularities, and situations where only the exterior solution exists.   

\subsection{A family of BH solutions}\label{sec_solns_1}

This family of solutions is parameterized by a function $h(\xi)$ via the relation \begin{equation}\label{hansatz}
M(f) = - \tfrac{f}{2} \left(1 + h(f^2)\right)\,,
\end{equation}
where we take
\be
\label{firstconditionsh}
h(0)=0\,,   \ \ \  h'(0) < \infty \,, 
\ee
conditions that allow making $f=0$  the far away region of the black hole. 
We also require that the correction to $M = -f/2$ implied by $h$ is `small'. For this we impose
\begin{equation}\label{small_condition}
|h(\xi)| \leq  1\,, \quad \ \forall \,   \xi \in \mathbb{R}\,.
\end{equation}
This implies, in particular, that $M(f) \leq  0$ for all $f > 0$. 
To get a horizon we need conditions on the behavior of $h$ for large argument:  
\be
\label{nodiv_conditions} 
\lim_{\xi \rightarrow \infty} h(\xi) = 0\,, \ \ \ 
\int_{0}^{\infty} h(\xi) d\xi  = \alpha < \infty\,,  
\ \ \ \   \lim_{\xi \rightarrow \infty} \xi h'(\xi)  = 0\,.
\ee	
The first condition 
implies that $M(f) \simeq -f/2$ for very large $f$.  The second 
condition strengthens
the bound $|h(\xi)|\leq  1$ by requiring that the full integral over positive arguments be finite.  The third condition follows from the first for a regular function at infinity. 

For the interior of the black hole, we need conditions on $h(\xi)$ for
large negative $\xi$.  At the cost of some generality, we simply demand that $h$ be 
an odd function of its argument:  
\begin{equation}\label{extra_condition}
h(-\xi) = - h(\xi)\,.
\end{equation}
The interior solution
demands one further constraint.  Let $f_0>0$ be the 
the value of $f$ where the function $\tilde P(f)$, defined in~\eq{Pf1}, 
first goes from negative to positive, so that
$\tilde P (f_0)=0$.   We will demand that 
\be
h(f_0^2) < 1\,,
\ee
in particular, $h(f^2)$ is not equal to one at this point. 

\medskip
\noindent
{\em \underline{Exterior solution:}} \ For the exterior solution, we have equations \eqref{fexte}. 
Picking the plus sign and inserting the function $M(f)$ as given in~\eqref{hansatz}, the equations for the metric and curvature reduce~to
\begin{subequations}
	\begin{align}   
	{m(f_2)\over m(f_1)} &=    \exp \biggl( \int_{f_1}^{f_2}  df I_{e}(f) ~\biggr)\,, 
	\quad I_{e}(f) \equiv \,  - \  \frac{1 + h(f^2)}  {2 \sqrt{P(f)} 
	} \,, \label{mh}\\
	R(f) &=  -\frac{f}{2}\left[ f (1 + h(f^2))^2 - 2 \left(1 + h(f^2) + 2 f^2 \frac{d h(f^2)}{df^2}\right)
	\sqrt{P(f)}
	\  \right] \label{Rh}\,,
	\end{align}
\end{subequations}
with the subscript $e$ for `exterior' and 
\begin{equation}
P(f) = 1 + \tfrac{1}{4} f^2 + \tfrac{1}{4} \int_0^{f^2} d\xi\, h(\xi)\,,
\end{equation}
which determines the location of branches. 
From the bound~\eqref{small_condition} on $h(\xi)$, we infer that
\begin{equation}\label{small_integral_condition}
- f^2 \leq  \int_0^{f^2} d\xi\, h(\xi) \leq  f^2\,,
\end{equation}
and we conclude that
\begin{equation}
P(f) \geq 1 \quad \forall\, f \in \mathbb{R}\,,
\end{equation}
showing that there are no branch cuts in the $f$ plane 
and we can work with 
\begin{equation}
f \in (0, \infty)\,.
\end{equation}
We now consider the behavior of the integrand $I_e$. 
Using the vanishing of $h(\xi=0)$ and the first two conditions in~\eqref{nodiv_conditions}, we conclude
\begin{equation}
\label{eorkdj0}
\lim_{f \rightarrow 0} I_{e}(f) = - \tfrac12\,, \quad \lim_{f \rightarrow \infty} I_{e}(f) \simeq  - \tfrac{1}{f} \quad \Rightarrow \quad 
\int_0^f d f' I_{e}(f') \simeq - \log f \,, \ \hbox{as} \ f \to \infty\, .
\end{equation}
For the ratio of metric values, these limits imply
\begin{equation}
{m(\infty)\over m(0)} =    \exp \biggl( \int_{0}^{\infty}  df I_{e}(f) ~\biggr)
= \exp(-\infty)  = 0\,,  
\end{equation}
which allows us to pick the integration constant such that
\begin{equation}\label{hm_ext_boundaries}
m(0) = 1\,, \quad m(\infty) = 0\,,
\end{equation}
the former consistent with $f=0$ being the faraway region and the 
latter consistent with $f=\infty$ being the horizon. 

Additionally, as required, the curvature goes to zero for $f \rightarrow 0$ and is finite for $f \rightarrow \infty$. While the former is a general feature of solutions coming from the classification, the behavior at the horizon is a consequence of \eqref{nodiv_conditions}, which imply
\begin{equation}\label{hR_ext_boundaries2}
\lim_{f \rightarrow \infty} R(f) \simeq - \frac{f^2}{2} \left[1 - \sqrt{1 + \frac{4 + \alpha}{f^2}}\right] \simeq  1 + \frac{\alpha}{4}\,.
\end{equation}

Let us now show that the behavior of the dilaton is as in the standard two-derivative
black hole:  the duality invariant dilaton diverges at the horizon but the scalar dilaton
is finite.  For this recall that 
\be
\Phi = \log f  + \Phi_1\,,
\ee
where $\Phi_1$ is a constant, 
and we can take the logarithm without concern as $f \in (0, \infty)$.  The asymptotically flat region is $f \sim 0$ and corresponds to $\Phi \to -\infty$ which is weak coupling. Using \eqref{ord-dil} and that
$m\simeq 1$ in this region, the scalar dilaton also indicates weak coupling.
The horizon is obtained as $f \to \infty$ and this means $\Phi \to \infty$ so we need to consider the metric contribution. 
We can obtain $\log m(f)$ by taking the log of \eqref{mh}, and picking the upper bound of the integration to be $f_2 = f$:
\begin{equation}\label{logm}
\log m(f) = \log m(f_1) +\int_{f_1}^{f} df' I_e(f') \,.  
\end{equation}
For large $f$, using the last relation in~\eq{eorkdj0} we find that
\be\label{estima2}
\log m \simeq -\log f   + c_2 \, , \ \ \  \hbox{as} \ \ f \to \infty \,. 
\ee
where $c_2$ is a constant. 
Therefore, at the horizon $x_h$, which corresponds to $f \to \infty$, from \eqref{ord-dil} we have
\be\label{phiathorizon}
\phi(x_h) \simeq  \tfrac{1}{2}  ( \log f  + \Phi_1 - \log f  + c_2)   \simeq c\,,
\ee
a constant, as we wanted to confirm.   

\medskip
\noindent
{\em \underline{Interior solution:}}  For the interior solution, we use~\eqref{tildeMM} to find that the ansatz \eqref{hansatz} gives 
\begin{equation}\label{Mhtilde}
\tilde M (\tilde f) =  -i M (- i \tilde f)  =  \tfrac{\tilde f}{2} \left(1 + h(- \tilde f^2)\right) = \tfrac{\tilde f}{2} \left(1 - h(\tilde f^2)\right)\,,
\end{equation}
where the last equality follows because $h$ is odd. The expressions for the metric ratios in~\eqref{fintm} and the curvature in~\eqref{curvxint}, become
\begin{subequations}\label{hmR_int}
	\begin{align}
	{\tilde m_{\pm}(\tilde f_2)\over \tilde m_{\pm}(\tilde f_1)} &=    \exp \biggl( \pm \int_{\tilde f_1}^{\tilde f_2}  df I_{i}(\tilde f) ~\biggr)\,, \quad I_{i}(\tilde f) \equiv \frac{1 - h(\tilde f^2)}{2\sqrt{\tilde P(\tilde f)}}\,, \label{mh_int}\\
	\tilde R_{\pm}(\tilde f) &=  \frac{\tilde f}{2}\left[ \tilde f (1 - h(\tilde f^2))^2 \pm 2\left(1 - h(\tilde f^2) - 2 \tilde f^2 \frac{d h(\tilde f^2)}{d\tilde f^2}\right)\sqrt{\tilde P(\tilde f)}\ \right] \label{Rh_int}\,,  
	\end{align}
\end{subequations}
with $i$ for `interior'. Regarding branch cuts, this time the range of $\tilde f$ depends on the roots of the function
\begin{equation}\label{tildeP}
\tilde P(\tilde f) = -1 + \tfrac{1}{4} \tilde f^2 - \tfrac{1}{4} \int_0^{\tilde f^2} d\xi\, h(\xi)\,.
\end{equation}
This function picks the following values at the boundaries
\begin{equation}\label{tildeP_limits}
\tilde P(0) = -1 < 0\,, \quad \lim_{\tilde f \rightarrow \pm \infty} \tilde P(\tilde f) 
= \lim_{\tilde f \rightarrow \pm \infty} \bigl(  -1 - \tfrac{1}{4} \alpha + \tfrac{1}{4} \tilde f^2\bigr)  = \infty \, ,
\end{equation}
where we used the second equation of \eqref{nodiv_conditions}. This 
shows that $\tilde P(\tilde f)$ must change sign as $\tilde f$ grows from zero. 
Let us call $f_0$ the value where $\tilde P(f_0) =0$; a branch point.  
Notice also that 
\be
\label{dertilde}
{d\tilde P\over d\tilde f} =   \tfrac{1}{2} \tilde f \, \bigl( 1 - h(\tilde f^2) \bigr)  \geq  0\,, 
\ee
since $|h| \leq 1$.  Note, however, that $h(f_0^2) < 1$ by assumption so that
\be
{d\tilde P\over d\tilde f} \Bigl|_{f_0}  > 0 \,.
\ee
Once $\tilde P(\tilde f)$ becomes positive it will remain
positive all the way since its derivative is never negative. 
Therefore, the range of validity of the solution is taken to be
\begin{equation}
\tilde f \in (f_0, \infty)\,.
\end{equation}
Using \eqref{small_integral_condition} one can show that $f_0 > \sqrt{2}$. For the standard BH we had $f_0 = 2$.   

We now investigate the condition for the horizon. We have
from~\eq{mh_int}  
\begin{equation}
\label{new03riu} 
{\tilde m_{\pm}(\infty)\over \tilde m_{\pm}(f_0)} =    \exp \biggl( \pm \int_{f_0}^{\infty}  d\tilde f I_{i}(\tilde f) ~\biggr) \,.
\end{equation}
We need the right hand side to be infinite for the plus sign; this makes $\tilde m(\infty)$ over the cut infinite, assuming $\tilde m(f_0)$ is finite. This is as desired.  For the minus sign the
right hand side should be zero, which makes $\tilde m(\infty)$ below the cut equal to zero --- this is as desired as it corresponds to the horizon.  To show this, we consider the integral of $I_i$ on the right-hand side.  At the branch point $f_0$ there is an integrable square-root singularity:  $\tilde P(f_0)=0$ and the derivative $\tilde P' (f_0)$ is nonzero, see \eqref{dertilde}.
We thus need only focus on the behavior of $I_i$ for large $\tilde f$.   
For the $\tilde f \rightarrow \infty $ limit, using \eqref{nodiv_conditions} we find
\begin{equation}
\label{lvaluesint}
I_{i}(\tilde f) \simeq  \frac{1}{\tilde f} \ \hbox{as} \ \ \tilde f\to \infty\, \quad \Rightarrow \ \quad  
\int_{f_0}^{\tilde f}  d \tilde f' I_{i}(\tilde f') \simeq  \ln \tilde f \,,  \ \  \ f \to \infty\,.
\end{equation}
Finally, inserting this into \eqref{new03riu} we get
${\tilde m_{\pm}(\infty)\over \tilde m_{\pm}(f_0)} =  
e^{\pm \infty}$, 
from where we can choose as integration constant a finite value of $\tilde m$ at the
branch point $\tilde m_+ (f_0) = \tilde m_-(f_0) = \tilde m_0$, so that we have, as desired
\be
\tilde m_{-}(\infty) = 0\,, \ \ \ \hbox{and} \ \ \ \tilde m_{+}(\infty) = \infty \,. 
\ee

The curvature at the branch point $f_0$  is easily evaluated since 
$\tilde P(f_0)$ vanishes: 
\begin{equation}\label{hR_int_boundaries1}
\tilde R_{-}(f_0) = \tilde R_{+}(f_0) = 2 \tilde M^2(f_0) = \tfrac{1}{2} f_0^2\left(1 - h(f_0^2)\right)^2\,. 
\end{equation}
For the infinite limit we have, from~\eqref{Rh_int}  
\be
\tilde R_\pm(\tilde f)  \simeq \tfrac{1}{2} {\tilde f^2} \Bigl[  1 \pm \sqrt{1 - \tfrac{4+\alpha}{ \tilde f^2}} \ \Bigr] \,, \ \ \  \tilde f \to \infty\,. 
\ee
As a result we find
\begin{equation}\label{hR_int_boundaries2}
\lim_{\tilde f \rightarrow \infty} \tilde R_-(\tilde f) = 1 + \tfrac{\alpha}{4}\,, \quad \lim_{\tilde f \rightarrow \infty} \tilde R_+(\tilde f) = \infty\,. 
\end{equation}
The first is the curvature at the horizon as computed from the interior. 
As required, it coincides with the curvature at the horizon computed from
the exterior in~\eqref{hR_ext_boundaries2}. The second result is the infinite curvature at the singularity.

For the duality-invariant dilaton in the interior region we have
\begin{equation}\label{Phitilde}
\tilde \Phi = \log \tilde f + \Phi_1\,,
\end{equation}
which is well defined for $\tilde f \in (f_0, \infty)$. $\tilde \Phi$ does not differentiate between lower or upper branches. It diverges at $\tilde f \to \infty$ but it is finite for $\tilde f = f_0$. For the scalar dilaton we use
\begin{equation}\label{ord-dilBHs}
\tilde \phi_{\pm}(\tilde f) = \tfrac{1}{2}\left(\tilde \Phi(\tilde f) + \log \tilde m_{\pm}(\tilde f)\right)\,,
\end{equation}
which does distinguish branches through $\tilde m_{\pm}$. While it is clear that $\tilde \phi_+(f_0) = \tilde \phi_-(f_0) = \text{finite}$, for $\tilde f \to \infty$ under and over the cut we need $\log \tilde m_{\pm}$. 
To this end, as we did for the exterior region, 
we integrate the metric ratio in~\eq{mh_int}   
\begin{equation}\label{logtildem}
\log \tilde m_\pm (f) = \log \tilde  m_\pm (f_1) \pm \int_{f_1}^{\tilde f} d\tilde f' I_i(\tilde f') \,.  
\end{equation}
Using large $\tilde f$ value of the integral from~\eq{lvaluesint}, we find
\be
\log \tilde m_\pm(f)  \simeq  \pm \log \tilde f   + c_\pm\,,\ \ \ \ \tilde f \to \infty \,, 
\ee
where $c_\pm$ are constants.
Plugging this expansion back into \eqref{ord-dilBHs}, together with \eqref{Phitilde} we find that for $\tilde f \to \infty$
\begin{equation}\label{phi-+infinity}
\tilde \phi_-(\tilde f) \simeq \tilde c\,, \quad \ \ \tilde \phi_+(\tilde f) \simeq \log \tilde f\,.
\end{equation}
The first one is consistent with the horizon interpretation, where we need $\tilde c = c$ with $c$ in \eqref{phiathorizon}, in order to match exterior and interior regions. The behavior of $\tilde \phi_+$ close to infinity signals, once more, the presence of a singularity there.

To sum up, the general ansatz \eqref{hansatz} parameterized with a function $h(\xi)$ satisfying \eqref{small_condition}, \eqref{nodiv_conditions} and \eqref{extra_condition} lead to BH solutions.  There are infinitely many functions
satisfying these conditions.  For example 
\be
h(\xi) = \xi \exp(-\xi^2)\,,
\ee
which satisfies $\int_0^\infty h(\xi) d\xi = \tfrac{1}{2}$ and 
$|h(\xi) | \leq {1\over \sqrt{2e}} < 1$.  

The way the more general 
solutions are parameterized in terms of $f$ mimics the standard BH. The exterior 
corresponds to $f \in (0, \infty)$ with $f=0$ the  faraway region and $f=\infty$ 
the horizon. The interior region is covered by two patches: $\tilde f \in (f_0, \infty)_{-}$ and $\tilde f \in (f_0, \infty)_{+}$, where the first goes under the branch cut (picking the minus sign in \eqref{hmR_int}) and the second over it (with the plus sign in \eqref{hmR_int}).
We begin under the branch at $\tilde f = \infty$ (the horizon) until $\tilde f = f_0$, the branch point, where all quantities are finite. From there, we
 return to $\tilde f = \infty$ 
over the branch cut. This time, the metric and curvature diverge at this point, which
is identified as the singularity.

\subsection{Cubic ansatz for $M(f)$}

The class of solutions introduced above assume a function
$h(\xi)$ that vanishes at zero and at infinity, among other conditions. 
Here we  assume that $h(\xi)$ is just
linear in $\xi$, leading to a {\em cubic} ansatz for $M(f)$:
\begin{equation}
M(f) = - \tfrac{1}{2} f + c\, f^3\,, \quad h(\xi) = - 2 c\,  \xi\,,
\end{equation}
with $c \neq 0$, otherwise we are back to the 
two-derivative BH. This $h(\xi)$ does not satisfy
the conditions in \eqref{small_condition} and \eqref{nodiv_conditions}, and so the physics is very different. 
We will find that the
solutions do not seem to have a BH interpretation due to the presence of naked singularities and disconnected regions.  This shows that arbitrary $\alpha'$ corrections can easily eliminate BH solutions.  
We were also motivated to consider this case  because some of the
pre-big bang solutions of~\cite{Gasperini:2023tus} were constructed with a cubic ansatz for $M(f)$.

Without going into the calculations which are straightforward,
 we proceed to enumerate the main characteristics of these solutions. To this end, we need the branch structures of exterior and interior solutions, which can be inferred respectively from
\begin{equation}
P(f) = 1 + \tfrac{1}{4}f^2 - \tfrac{c}{4}f^4\,, \quad \hbox{and}
\quad  \tilde P(\tilde f) = - 1 + \tfrac{1}{4}\tilde f^2 + \tfrac{c}{4}\tilde f^4\,. 
\end{equation}
The analysis is different depending on the sign of $c$. For simplicity we give the results for $f > 0$, but similar results hold for the negative region, which is connected to the positive one via T-duality.

\begin{itemize}
	
	\item $c> 0$:
	
	The exterior region is parameterized by 
	$f \in (0, f_0)$, where
	\begin{equation}\label{f0cubic}  
	f_0^2 = \frac{\sqrt{1 + 16 c} + 1}{2 c}\simeq \frac{1}{c} + \mathcal{O}(\sqrt{c})\,,
	\ \  c \to 0 \,.
	\end{equation}
	For positive $f$, the function $P(f)$ only vanishes at $f_0$.
	Since $f_0\sim 1/\sqrt{c}$, the appearance of 
	a branch cut for the exterior solution is non-perturbative in $c$.
	 As it should be, $f=0$ still corresponds to the far away region. 
	 This time, the boundary at $f_0$ is a point with non-vanishing metric 
	 and {\em finite} curvature.
	
	The interior region exists for $|\tilde f| \geq \tilde f_0$,    
	with $\tilde f_0^2 = \tfrac{\sqrt{1 + 16 c} - 1}{2 c}$.  
	As in the standard BH and the class of deformations studied above, we cover the solution by the two patches $\tilde f \in (\tilde f_0, \infty)_- \cup (\tilde f_0, \infty)_+$, corresponding to going under the cut and then coming back over the cut. At one end, 
	$\tilde m_{-}=0$ but $\tilde R_{-} = \infty$.  At the other end 
	$\tilde m_{+} = \infty$ and $\tilde R_{+} = \infty$. 
	
	The absence of a horizon in the exterior and the presence of singularities in the interior makes this solution incompatible with a black hole.  Moreover, the two geometries do not connect.  Each one is a separate solution.  
	
	\item $c < 0$:
	
	In this situation, $P(f)$ is always positive and therefore the exterior solution is valid for $f \in (0, \infty)$ as in the two-derivative case. 
	As usual, $f\sim 0$ is the AF region,
	but $f \to \infty$ is not a horizon but a point with
	 vanishing metric with infinite curvature. This region then corresponds to a naked singularity.
	
	The interior region has now two branch points, $\tilde f_0$ and $\tilde f_1$. The point $\tilde f_0$ is the same as for the $c>0$ case, and $\tilde f_1$ is
	given from $\tilde f_1^2 = \tfrac{-\sqrt{1 + 16 c} - 1}{2 c}$. The interior region only exists for $c > - \tfrac{1}{16}$ and $\tilde f \in (\tilde f_0, \tilde f_1)$. Since this $\tilde f$ interval is finite, is easy to see that $\tilde m_{\pm}$ and $\tilde R_{\pm}$ 
	are regular everywhere but the metric never vanishes.   For $c<- \tfrac{1}{16}$, there is no interior solution, just an exterior one.

	When both exist, the exterior and the interior solutions do not combine.  They are separate solutions, and we have no black hole.\footnote{The construction 
	of~\cite{Ying:2022xaj} was in this class of negative $c$ theories, so while there seems to be a regular interior solution, the exterior solution does not work out.}

\end{itemize}

\section{Horizons and singularities} \label{horandsing}

Having gained experience on the constraints of black hole solutions in
the previous section, we now want to discuss some general lessons.
We show that for dilaton-gravity theories in two-dimensional time-independent
backgrounds  
 T-duality implies a constraint on
the construction of regular black holes (geometries
with a horizon but no curvature singularity).   We will prove that T duality
maps a geometry with a regular horizon to one with a curvature singularity.
This  shows that the observation of Giveon~\cite{Giveon:1991sy} for the 
two-derivative black hole is
general. 
For a solution that is expected to be self-dual, this means one cannot
have a horizon and no singularity.   
Moreover, we explain,
as a corollary, how the usual branch structure of the interior solution
implies self-duality and thus a singularity. 
In the next section we will give black hole solutions that have an unusual 
branch structure and that are \textit{not} self-dual so that the singularity 
\textit{can} be avoided.

To prove that a horizon implies a singularity in the dual geometry we begin
with a configuration where,  without loss of generality, we can take 
$x=0$ as the position of the horizon, the point where the metric $m(x)$ vanishes.  
A regular configuration is one where the scalar curvature is finite everywhere. 
Since the conclusions should be independent of any gauge choice, we pick $n=1$ for simplicity.

Demanding $R= - 2m''/m$ (\eqref{scalarR}) to be finite everywhere implies, in particular, ${m''(x)}/{m(x)}$ must be finite at the horizon. Then, from $\hat R = - R - 4 M^2$,
 (\eqref{dualR}), we see that regularity of the dual curvature $\hat R$ additionally forces $M= m'/m$ to be finite everywhere, in particular at the horizon. Together with the horizon condition we obtain
\be
\label{rtildrcond}
\lim_{x \to 0} m(x) = 0 \,, \ \ \ 
\lim_{x \to 0}  {m''(x) \over m(x) } = \hbox{finite} \,,  \ \ \ \ \ \ 
\lim_{x \to 0}  {m'(x) \over m(x) } =  \hbox{finite} \,.
\ee

To gain intuition, we now show that \eqref{rtildrcond} cannot be satisfied for analytic metric components. If $m(x)$ is analytic, it admits a regular 
Taylor expansion around $x=0$, the location of the horizon,
\be\label{mexpansion}
m = a_1  x  +  \tfrac{1}{2}  a_2 x^2 +  \tfrac{1}{3} a_3 x^3  +  \cdots
\ee
where the first condition in \eqref{rtildrcond} rules out the presence of a constant term. Plugging \eqref{mexpansion} into the second condition of \eqref{rtildrcond} we find
\begin{equation}
\lim_{x \to 0}  {m''(x) \over m(x) } \simeq \frac{a_2}{a_1} \frac{1}{x} + \dots
\end{equation}
which is finite only if $a_2=0$. Finally, with $a_2=0$ the third condition on \eqref{rtildrcond} now reads
\begin{equation}
\lim_{x \to 0}  {m'(x) \over m(x) } \simeq \frac{1}{x} + \dots
\end{equation}
which diverges regardless the coefficients of the Taylor expansion.  This logarithmic
derivative diverges at the position of the horizon.   As claimed the conditions cannot be satisfied.  

\noindent
{\bf A proof that a horizon implies a curvature singularity}

\noindent
Having considered the case when
$m(x)$ is analytic we now build a general proof.  
We define the neighborhood $N_\epsilon= (0,\epsilon)$, 
and the neigborhood $\bar N_\epsilon= [0, \epsilon)$ where
we include the point $0$.
We assume that there is an $\epsilon$
sufficiently small such that the following hold:

\begin{enumerate} 

\item
The function $m(x)$ is continuous in $\bar N_\epsilon$ 
and vanishing at $x=0$. This is 
an isolated zero: the neighborhood $\bar N_\epsilon$ does not contain another zero
of $m$.

\item The curvature $R(x)$  in $\bar N_\epsilon$ is finite. 

\end{enumerate} 
These are the natural conditions satisfied by a conventional horizon. 

\noindent
{\bf Claim:} {\em  The dual curvature $\hat R$ 
cannot be finite in $\bar N_\epsilon$.}

\noindent
{\em Proof.} 
Since the curvature $R$ and the dual curvature $\hat R$ are related as follows
\be
\hat R = - R - 4 \Bigl( {m'\over m} \Bigr)^2 \,,
\ee
and $R$ is assumed finite in $\bar N_\epsilon$ this means
that the finiteness condition
\be
\label{contradictionass}
{m'(x)\over m(x) } <  \infty \,, \ \  \ \  \  \forall x \in \bar N_\epsilon\,,
\ee
cannot hold.  Let us assume it holds and derive a contradiction.

Since the sign of $m(x)$ is immaterial to the 
definition of the metric component $g_{00}$, we can change it at will
so that we can require
$m(x) > 0$ for  $x\in N_\epsilon$.  This follows from the
continuity of $m$ and the isolated zero.
Given the positivity of $m(x)$ in $\bar N_\epsilon$ we can write $m(x)$ 
in terms of its logarithm $g(x)$: 
\be
m (x)= e^{g(x)} \,,   \ \ \hbox{with} \ \ m(0) = 0 \quad \to \quad g(0)= -\infty \,.  
\ee
Since $m(x) \not=0$ in $N_\epsilon$, the function $g(x)$ is finite in $N_\epsilon$.
Taking a derivative of $m$ we have 
\be
{m'(x)\over m(x)} = g'(x)  \,. \   
\ee
Given the hypothesis~\eqref{contradictionass} we conclude that
$g'(x) < \infty$ in $\bar N_\epsilon$. 
Now consider the relation\footnote{We thank Daniel Harlow for a discussion
where he suggested this line of reasoning.} 
 \be
g(0) =  g(x) - \int_0^x g'(u) du\, ,\ \ \ \forall \ x \in N_\epsilon\,, 
\ee
and examine the right-hand side above. 
For $x\in N_\epsilon$,  the function $g(x)$ is finite.
Moreover, since $g'$ is finite in $\bar N_\epsilon$ and $x$ is finite,
the integral is also finite.  But with both terms on the right-hand side
finite, we cannot have  $g(0)= -\infty$ as required.  
This contradiction means that~\eqref{contradictionass} does not hold;
we need $g'$ to diverge in order for the integral above to give an infinite $g(0)$.
This is what we wanted to show.   The proof does not tell us where is
$\hat R$ infinite, but the intuition of the analytic case indicates
that it is expected to be at $x=0$.  \hfill $\square$

\medskip
The result above is reasonable: In one dimension, T-duality maps $m(x) \rightarrow {1}/{m(x)}$, which makes a vanishing metric unavoidably dual to a divergent one. 
All we had to prove was that such dual metric has divergent curvature. 

This result actually gives a simple proof that the interior solution based
on the contour extending below and above the branch cut in $f$ space
must lead to a singularity.  The main observation here is simple.  Recall that
$\tilde R$ is given in~\eq{Rintx}
and its dual $\hat{\tilde R}$ is obtained by letting $\tilde M \to -\tilde M$, 
\be
\begin{split}
\tilde R \ &  =  \ 2 \tilde M^2  +  2 D\tilde M\,,  \\
\hat {\tilde R} \   &= \  2 \tilde M^2  - 2 D\tilde M \,. 
\end{split}
\ee 
Compare with equation~\eq{curvxint} for the curvatures above ($+$) and below
($-$) the cut: 
\be\label{Rtilde+-} 
\tilde R_{\pm}(\tilde f) =    2 \tilde M^2 \pm   2 \tilde f \, {d\tilde M\over d\tilde f}  
\sqrt{ \tilde P(\tilde f)  } \ \,. \ee
Since $\tilde M(\tilde f)$ and $\tilde P(\tilde f)$ are functions
without branches, each term in the above curvature formula takes the same value above or below the cut;  the only
difference being that they enter with different sign combinations for the curvature
and for its dual.  It thus follows that we can identify $\tilde R = \tilde R_+$ and
$\hat{\tilde R} = \tilde R_-$.  The curvatures above and below the cut are related
by T-duality!  Since the $\tilde f$
contour defining the $x$ domain extends both above and
below the cut, this means that having a horizon below the cut (as we conventionally
set it up) will imply a curvature singularity above the cut.   The interior solution is
self-dual:  T-duality, supplemented with the time reparameterization that
exchanges the values of $x(\tilde f)$ for points immediately above and below the cut,
leaves the fields invariant.

\section{Regular black holes} \label{resthebhsing}

In this section we modify the analysis of  section~\ref{sec_solns_1} where we found
black hole solutions with conventional properties. We will show that with a couple
of extra constraints on the function $M(f)$ we can obtain 
{\em regular} black holes:  solutions where the exterior, as usual, has an asymptotically
flat region and a horizon, and the interior has no singularity in the metric, 
the curvature, or the dilaton. As in section~\ref{sec_solns_1}  we parameterize $M$ in terms of a function~$h$
\begin{equation}\label{Mfh}
M(f) = - \tfrac{1}{2}f \left[1 + h(f^2)\right]\,,
\end{equation}
with $h(\xi)$ satisfying the same properties as before: $h(0)=0$, $h'(0)$ is finite, 
$|h(\xi)| \leq 1$ for all $\xi$, $h(\infty) = 0$,  $\xi h'(\xi)|_{\xi =\infty} = 0$,  the parity condition $h(-\xi) = -h(\xi)$, and $\int_0^\infty h(\xi) d\xi = \alpha < \infty$. 
If this the case, the interior solution is parameterized by 
\begin{equation}\label{Mfhtilde}
\tilde M(\tilde f) = \tfrac{1}{2} \tilde f\left[1 - h(\tilde f^2)\right]\,.
\end{equation}
There was one additional condition on $h(\xi)$ (recall that $\xi= f^2$): it is not supposed to reach the value of 
one at the point $f_0 > 0$ where the function $\tilde P(f)$ first goes from negative to positive.  
We will explicitly relax this 
condition; this will be essential for the interior
solution.  Letting $\xi_0 = f_0^2$, we demand that 
\begin{equation}\label{xi0_conditions}
h(\xi_0) = 1\,, \ \ \ h'(\xi_0) = 0\,, 
\quad h''(\xi_0) < 0\,, \quad \hbox{and}
\quad \int_0^{\xi_0} d\xi\, h(\xi) = \xi_0 - 4\,,
\end{equation}
We will confirm that with these conditions $\tilde P(f_0)=0$.  This fact and the motivation
for the above set of conditions will be explained below.   

If we call ${\cal T}$ the space of all possible $\alpha'$ corrections (the space of
all $M(f)$'s), and ${\cal S}$ the subspace of section~\ref{sec_solns_1} leading to
conventional 2D black holes, the regular black holes arise from the boundary
of ${\cal S}$.  This is clear because the inequality $h(f_0^2) < 1$, needed for
conventional black holes, is saturated for regular black holes. 
They do not cover the full boundary,
because we require additional conditions.

We will now examine the equations of motion in the $f$-parameterization, both
in the exterior and the interior, finding solutions.   We will see that in the 
interior the non-invertibility of $\tilde M(\tilde f)$ plays a central role.
We will also explain that the solution for the regular black hole has
an interior that is a cosmology that at late times is  asymptotic to Minkowski space with a constant dilaton.  
This behavior suggests strongly
that this is not a solution in string theory.

\subsection{Regular BH solutions} \label{regbhsoljnvg}

\noindent
{\em \underline{Exterior solution:}}
The additional conditions~\eq{xi0_conditions} preserve all good features of the exterior BH solution. In particular, for the exterior region $f_0$ has no special meaning, it is just a finite intermediate point in the solution's domain. The object that determines the latter is the argument of the square root: 
\begin{equation}
\label{pfgovmbvg}
P(f) = 1 + \tfrac{1}{4} {f^2}+ \tfrac{1}{4} \int_0^{f^2} d \xi\, h(\xi)\,,
\end{equation}
and the domain $f\in (0,\infty)$ that we use remains unchanged
since, as a consequence of the bound $|h(\xi)| \leq 1$, 
we have $P(f) >0$ for such domain.   The analysis of the exterior
solution in section~\ref{sec_solns_1}
remains unchanged, and the results for the metric, curvature,
and dilaton derived there hold here as well.  

\medskip
\noindent
{\em \underline{Interior solution:}} 
For the interior solution, the role of $f_0$ is crucial to get a regular solution.
The  idea is that the conditions~\eqref{xi0_conditions} render  the position $x(f_0)$   
infinitely faraway. Therefore, after going from the horizon ($\tilde f = \infty$) to $\tilde f = f_0$ under the cut, we {\em do not need} to go back to $\tilde f=\infty$ on the top of the cut,   
where the singularity always was.  
Now, we can just stop at $f_0$ because in terms of $x$ this is the end of space-time.

We begin with the function $\tilde P (\tilde f)$ appearing in the interior formulae
under a square root:
\begin{equation}\label{Pftilde}
\tilde P(\tilde f) =- 1 + \int_0^{\tilde f} \tilde M (\tilde f') d\tilde f'  =   -1 + \tfrac{1}{4}\tilde f^2 - \tfrac{1}{4} \int_0^{\tilde f^2} d \xi\, h(\xi)\,.
\end{equation}
We then have
\begin{equation}
\label{smlimits}
\tilde P(\tilde f=0) = -1\,, \quad \tilde P(f_0) = 0\,, \quad \tilde P(\tilde f) \simeq \tfrac{1}{4} 
{\tilde f^2} \ \ {\rm for} \ \  \tilde f \to \infty\,. 
\end{equation}
The first equality is manifest, the second equality follows from the last condition in \eqref{xi0_conditions}, and the last one from the convergence of $\int_0^\infty h(\xi) d\xi$. To understand the nature of the interior solution we must consider derivatives of
$\tilde P(\tilde f)$. The first derivative  coincides with $\tilde M(\tilde f)$: 
\begin{equation}\label{M_positive}
\tilde P'(\tilde f) = \tilde M(\tilde f) = \tfrac{\tilde f}{2} \left[1 - h(\tilde f^2)\right] \geq 0 \quad \forall\, \tilde f \geq 0\,,
\end{equation}
where the inequality follows from~$|h|\leq 1$. We then have the following values
\begin{equation}
\tilde P'(0) = \tilde M(0) = 0\,, \quad \tilde P'(f_0) = \tilde M(f_0) = 0\,, \quad \tilde P'(\tilde f) = \tilde M(\tilde f) \simeq \tfrac{1}{2} {\tilde f} \ \ {\rm for} \ \ \tilde f \to \infty\,.
\end{equation}
The first equality is manifest, the second follows from $h(\xi_0) =1$, the third
from the vanishing of $h$ for large argument. The second derivative is given by
\begin{equation}
\tilde P''(\tilde f) = \tilde M'(\tilde f) 
=  \tfrac{1}{2}\bigl( 1 - h(\xi)\bigr) - \xi \, h'(\xi)\,,
\end{equation}
using $\xi = \tilde f^2$.  We obtain  
\begin{equation}
\tilde P''(\tilde f=0) = \tilde M'(\tilde f=0) 
= \tfrac{1}{2}\,, \quad \tilde P''(f_0) = \tilde M'(f_0) = 0\,.
\end{equation}
The second equality holds because $h(\xi_0)=1$ and $h'(\xi_0) = 0$. 
For the third derivative we have
\begin{equation}
\tilde P'''(\tilde f) = \tilde M''(\tilde f) =- 3 \tilde f  h'(\xi) - 2 \tilde f^3 h''(\xi) \,.
\end{equation}
The special values here are
\be
\label{spvaluesMf0}
\tilde P'''(\tilde f = 0) = \tilde M''(\tilde f= 0) = 0\,, \quad \tilde P'''(f_0) = \tilde M''(f_0) = -2 f_0^3\, h''(\xi_0) > 0\,,
\ee
the second relation following from the assumption $h''(\xi_0) < 0$. 

All these results can be represented collectively in the two plots in 
Fig.\ref{figBHext_MP_plots}, 
where
we show $\tilde M (\tilde f)$ and $\tilde P(\tilde f)$.

\begin{figure}[ht]
	\centering
\epsfysize=6.0cm
	\epsfbox{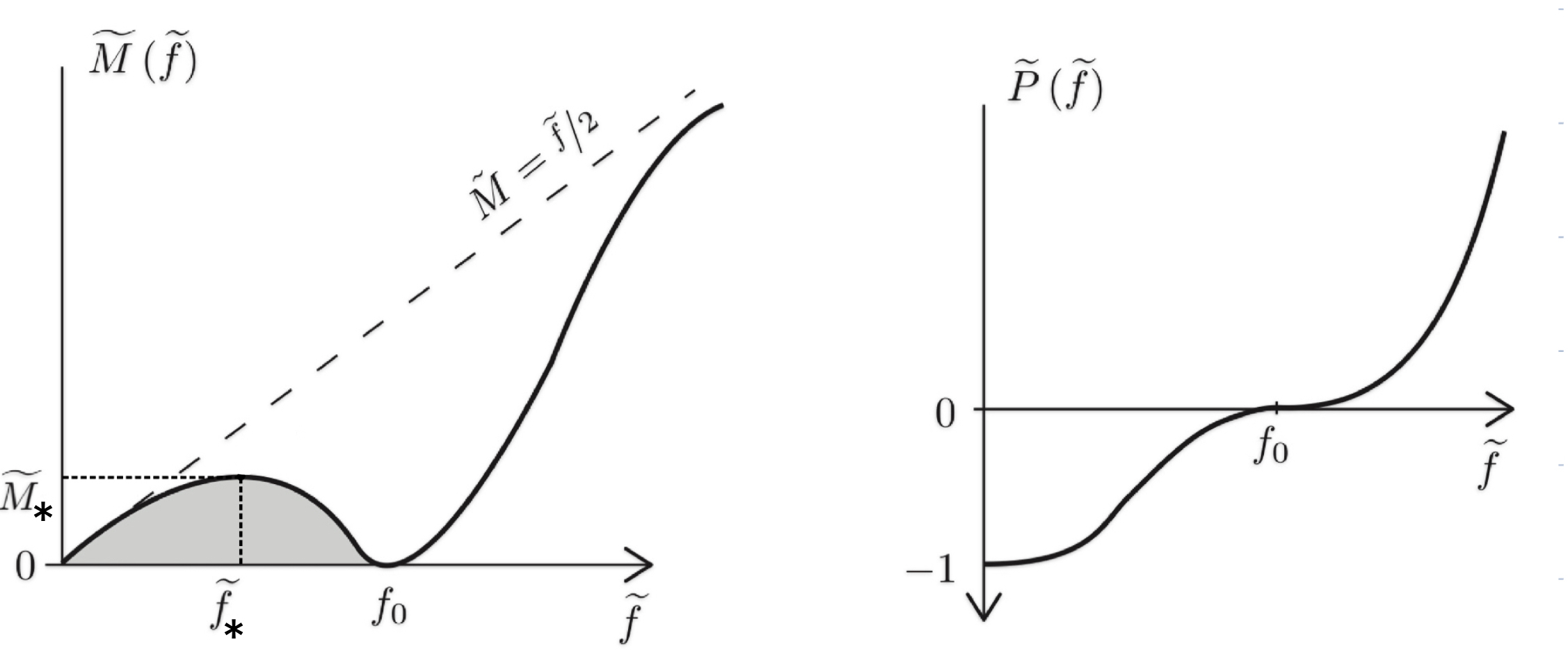}
	\caption{\small Left: Sketch of $\tilde M(\tilde f)$.  The shaded region 
	must have unit area, 
	the point $\tilde f_*$ indicates the position of the first local maximum $\tilde M(\tilde f_*) = \tilde M_*$, 
	the point $f_0$ is a minimum, and for large $\tilde f$ the curve approaches the line $\tilde f/2$.
	Right: Sketch of  $\tilde P (\tilde f)$.  This function 
	 changes sign at $f_0$, which is an inflection point.}
	\label{figBHext_MP_plots}
\end{figure}

The plot of $\tilde M$ (left) shows a function that begins at zero with positive slope and remains strictly positive until $f_0$ where $\tilde M(f_0) = \tilde M'(f_0) = 0$ and $M''(f_0) > 0$, which makes $f_0$  
a minimum. The area under the curve in the
interval $(0, f_0)$ is exactly one because
$\tilde P(f_0) = -1 + \int_0^{f_0} d \tilde f \tilde M(\tilde f) = 0.$
For $\tilde f > f_0$ the function remains always positive (see \eqref{M_positive}) and approaches infinity exactly as the standard BH. 

Let us now turn to the graph of $\tilde P(\tilde f)$. This function equals $-1$ for $\tilde f=0$, 
the minimum of $\tilde P(\tilde f)$. The function then increases with $\tilde f$ until $f_0$, where it vanishes together with its first and second derivative. Since the third derivative is nonzero, in fact, $\tilde P'''(f_0) > 0$, we see that
$f_0$ is an inflection point. Afterwards, $\tilde P(\tilde f)$ keeps increasing monotonically, as can be seen from \eqref{M_positive}, and at infinity it behaves as ${\tilde f^2}/{4}$.

Note now that with $h(\xi)$ as discussed, the domain for the interior solution is $\tilde f \in (f_0, \infty)$.
With the function $\tilde P(\tilde f)$, its first derivative,  and its second derivative all vanishing
at $f_0$, we see that
$\tilde P$ near $f_0$ is given by 
\begin{equation}\label{P3expansion}
\tilde P(\tilde f) \simeq  \ \tfrac{1}{6} \tilde P'''(f_0) (\tilde f - f_0)^3 \ =\ -\tfrac{1}{3} f_0^3 h''(f_0^2)(\tilde f - f_0)^3 > 0\quad 
 \ \ {\rm for} \ \
 \tilde f \sim f_0 \,. 
\end{equation}
This behavior will allow us to regularize the BH's interior.

We first confirm the presence of the horizon in this interior solution (below the cut).
Consider the third equation in~\eq{fint},    
picking the minus-sign and integrating over the whole range of $\tilde f$:
\begin{equation}\label{m0infty}
\frac{\tilde m(\infty)}{\tilde m(f_0)} = \exp\Biggl(- \int_{f_0}^{\infty} d \tilde f 
\ {1 - h(\tilde f^2) \over 2 \sqrt{\tilde P(\tilde f)}}  \Biggr) \,. 
\end{equation}
The integral is convergent at the lower limit since the integrand  in fact vanishes
there (the numerator vanishes like $(\tilde f-f_0)^2$ and the denominator vanishes
like $(\tilde f-f_0)^{3/2}$.  At the upper limit we get a logarithmic 
divergence since $\sqrt{\tilde P} \sim  \tilde f$ (see~\eq{smlimits}) and the numerator approaches a constant.  
Thus the integral is infinite 
and the right-hand side of \eqref{m0infty} becomes zero. 
Since $\tilde m(f_0)$ is finite we can choose
\begin{equation}
\tilde m(f_0) = 1\,, \quad  \tilde m(\infty) = 0\,.
\end{equation}
This is the horizon at $\tilde f = \infty$. 

For the curvature, we can use~\eqref{Rtilde+-}
with the lower sign, as we
are working below the cut, 
\begin{equation}
\tilde R_-(\tilde f)
 =  2 \Bigl[(\tilde P'(\tilde f))^2 - \tilde f \tilde P''(\tilde f) \sqrt{\tilde P(\tilde f)} \ \Bigr]\,,
\end{equation}
where we rewrote the dependence on $\tilde M$ in terms of derivatives of $\tilde P$. The curvature  vanishes at $\tilde f = f_0$, because $\tilde P, \tilde P',$ and 
$\tilde P''$ all vanish at $f_0$. For $\tilde f \rightarrow \infty$ the behavior of $\tilde P$ is
the same as we had in section~\ref{sec_solns_1}, and therefore
the curvature computation we did there applies here as well.  The relevant 
result is given
by the first expression in~\eq{hR_int_boundaries2}.   In summary, we have
\begin{equation}
\tilde R_{-}(f_0) = 0\,, \quad \tilde R_{-}(\infty) = 1 + \tfrac{\alpha}{4}\,. 
\end{equation}

For the dilaton we have, as usual,  $\tilde\Phi(\tilde f) = \log \tilde f   + \Phi_1 $, which is finite for $f_0$ and diverges for $\tilde f \to \infty$. The scalar dilaton is given by $\tilde \phi = \tfrac{1}{2} ( \tilde\Phi + \log \tilde m)$. For $\tilde f=f_0$ we get $\tilde \phi (f_0) =  \tfrac{1}{2} \tilde\Phi(f_0)$  since $\tilde m (f_0) =1$.
The scalar dilaton, as opposed to the duality invariant dilaton $\tilde\Phi$, has no divergence as $\tilde f\to \infty$.  The argument follows identically as the one used in section \ref{sec_solns_1} to arrive at \eqref{phi-+infinity}, which in this case reads
\begin{equation*}
\tilde \phi(\tilde f) \simeq \tilde c \ \ {\rm for} \ \ \tilde f \to \infty\,.
\end{equation*}

Finally, the relation between $x$ and $\tilde f$ is very different from that
in the family of singular black holes:  the point half-way from the horizon 
to the singularity
is moved to infinite distance.  This is the branch point at $f_0$.
Using the behavior~\eqref{P3expansion} of
$\tilde P(\tilde f)$ close to $f_0$, the differential equation \eqref{x-diffe} relating
$x$ and $\tilde f$ takes the form
\begin{equation}\label{dxdfregular}
dx = - d \tilde f \frac{1}{\tilde f \sqrt{\tilde P(\tilde f)}} \simeq - d \tilde f \sqrt{\frac{3}{f_0^5 |h''(\xi_0)|}} \ (\tilde f - f_0)^{-\tfrac{3}{2}} 
\ \ {\rm for} \ \  \tilde f \sim f_0 \,. 
\end{equation}
This integral diverges as $\tilde f\to f_0$. On the other hand, for $\tilde f \rightarrow  \infty$ we get a finite integral: here $\tilde P (\tilde f) \simeq \tfrac{1}{2} \tilde f$ and therefore
$dx \simeq  -2{d\tilde f}/\tilde f^2$, which is integrable as $\tilde f \to \infty$. This allows us to define $x(\tilde f)$ with the following values at the boundaries
\begin{equation}
x(f_0) = \infty\,, \quad x(\infty) = 0\,.
\end{equation}

These results make the interior region to be a regular cosmology, 
identifying $x$ with a time coordinate
$\tau$ and $t$ with a spatial coordinate $w$, as required by the signature of the metric
\be
ds^2 =  \tilde m^2(x) dt^2  - dx^2   =  - d\tau^2  + \tilde m^2 (\tau) dw^2  \,. 
\ee 
This cosmology begins at time $\tau=0$ with finite curvature $\tilde R(\tau=0) = 1 + \tfrac{\alpha}{4}$ 
and evolves as  $\tau\rightarrow \infty$
to a geometry with zero curvature and hence to flat space. 
In this cosmology there is no big-bang, and time begins at the position of the horizon. 

We now perform a local analysis to see how the solution approaches the flat region to compare it with standard cosmology. For clarity we continue to use $x$ (rather than
$\tau$). We begin by integrating \eqref{dxdfregular} to get the relation between $x$ and $\tilde f$ close to $f_0$
\begin{equation}
x(\tilde f) \simeq \sqrt{\frac{12}{f_0^5 |h''(\xi_0)|}}\  {1\over \sqrt{\tilde f - f_0}} \ \ {\rm for} \ \
\tilde f \sim f_0 \,, 
\end{equation}
where we dropped an integration constant. Inverting this expression we get
\begin{equation}
\tilde f(x) \simeq  f_0 + \frac{12}{f_0^5 |h''(\xi_0)|}\, \frac{1}{x^2}\ \ {\rm for} \ \  x \to \infty \,.
\end{equation}
With this relation we can express all quantities locally in terms of $x$. For $\tilde M(x)$ we use \eqref{Mfhtilde} to expand $\tilde M (\tilde f)$ around $\tilde f \sim f_0$:
\be
\tilde M (\tilde f)  = \tilde M (\tilde f_0) + \tilde M'(f_0) ( \tilde f - f_0) 
+ \tfrac{1}{2}  \tilde M''(f_0) ( \tilde f - f_0)^2 + {\cal O} ((\tilde f - f_0)^3) \,. 
\ee
Recalling that $\tilde M(f_0) = \tilde M'(f_0) = 0$ and using the value
for the second derivative from~\eq{spvaluesMf0} we get
\begin{equation}
\tilde M(\tilde f) \simeq\   f_0^3|h''(\xi_0)|\ (\tilde f-f_0)^2\ \ {\rm for} \ \ \tilde f \sim f_0\,, 
\end{equation}
which in terms of $x$ reads
\begin{equation}
\tilde M(x) \simeq \frac{144}{f_0^7 |h''(\xi_0)|}\, \frac{1}{x^4}\ \ {\rm for} \ \  x \to \infty \,. 
\end{equation}
Now $\tilde m(x)$ is given by
\begin{equation}
\tilde m(x) = C\exp\left(\int^x dx' \tilde M(x')\right) \simeq C \exp\left(- \frac{48}{f_0^7 |h''(\xi_0)|}\, \frac{1}{x^3}\right) \simeq C \Bigl(  1 - \frac{48}{f_0^7 |h''(\xi_0)|}\, \frac{1}{x^3}\Bigr) \,,
\end{equation}
with $C$ a constant of integration. 
For the curvature we use \eqref{Rint} to get
\begin{equation}
\tilde R(x) = 2\left(\tilde M^2(x) + \tilde M'(x)\right) \simeq 2 \tilde M'(x) \simeq -\frac{1152}{f_0^7 |h''(\xi_0)|}\, \frac{1}{x^5}\,, 
\end{equation}
where we noted that $\tilde M^2 \sim x^{-8}$ and thus can be ignored to lowest order. 
Finally, the duality invariant dilaton is given by 
\be
\tilde \Phi = \log \tilde f + \Phi_1  \simeq  \tfrac{1}{f_0} (\tilde f - f_0)  + \log f_0 + \Phi_1 \ \ {\rm for} \ \  \tilde f \sim f_0 \,,
\ee
which in terms of $x$ reads
\begin{equation}
\tilde \Phi(x) \simeq \ \frac{12}{f_0^6 |h''(\xi_0)|}\, \frac{1}{x^2}   + \log f_0 + \Phi_1  \,. 
\end{equation}
The scalar dilaton is 
given by $\tilde \phi(x) \simeq \tfrac{1}{2} \tilde \Phi(x)$ 
since the contribution from the logarithm of $\tilde m$ is of order $x^{-3}$ and thus subleading. 

\noindent   
{\underline{Remark:}}  It is important to note that the interior solution, for infinite time $x$,  becomes flat Minkowski space with a {\em constant} dilaton. This may seem surprising, given the well-known fact that the `cosmological' term in the 
2D action requires a linear dilaton when the spacetime becomes flat. 
Indeed, the black hole exterior in the far away region operates in this way, both in the two-derivative theory and in the case
above: the metric approaches the flat metric and the dilaton profile approaches a linear function.  To see what is happening in the interior consider the exterior and interior relations~\eq{eorid38}, \eq{dPhiout}, and~\eq{eorid39},~\eq{fintP}: 
\be
\begin{split}
\label{gcomgzbvg}
 \Bigl( {d\Phi\over dx} \Bigr)^2 = & \ \ \ \  1 -  g ( f)  \ =    P(f) \,,\\
 \Bigl( {d\tilde \Phi\over dx} \Bigr)^2 = & \  -1 + \tilde g (\tilde f)  =   \tilde P(\tilde f) \,.
\end{split} 
\ee
Moreover, 
 just 
 as we have~\eq{extentsiong} in the exterior,
 there is a similar equation in the interior
\be
\label{gcomclbvg}
\begin{split}
\check g (M)  \ &= \   g(f (M))   \, , \qquad M \in [0, M_*]\,, \\
\tilde {\check g} (\tilde M) \ &= \ \tilde g ( \tilde f (\tilde M))\,, \qquad  \tilde M \in [0, \tilde M_*] \,,
\end{split}\ 
\ee
where $\tilde M_*$ 
is the value beyond which the inverse $\tilde f (\tilde M)$ of $\tilde M(\tilde f)$ is not expected to have a convergent series. 
Looking at Fig.\,\ref{figBHext_MP_plots}, $\tilde M_*$ is the maximum of $\tilde M(\tilde f)$ attained for  $\tilde f_* \in (0, f_0)$. 
The non-invertibility issue can be better appreciated in Fig.\,\ref{fig_multivalue}.

\begin{figure}[ht]
	\centering
	\epsfysize=6.0cm
	\epsfbox{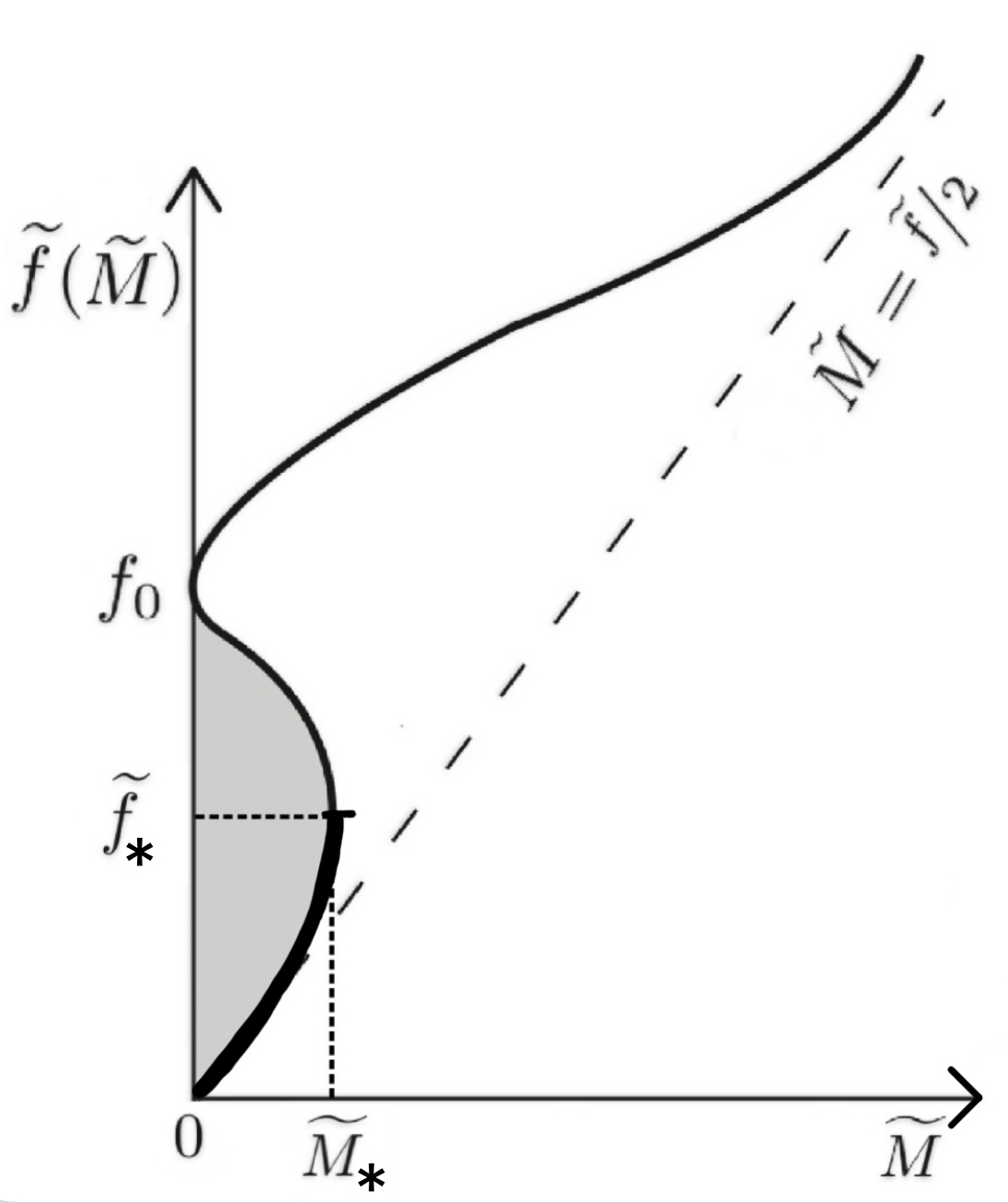}
	\caption{\small Sketch of $\tilde f(\tilde M)$, the inverse of $\tilde M (\tilde f)$ in Fig.\,\ref{figBHext_MP_plots}, making its multi-valuedness evident. 
The bold curve is $\tilde f (\tilde M)$ on the domain of expected
convergence of its series representation.}
	\label{fig_multivalue}
\end{figure}

Consider first the exterior, in particular the AF region $f\sim 0$ and $M \sim 0$. Here $P(f=0) = 1$ (see~\eq{pfgovmbvg}), and therefore $g(0)=0$ (\eq{gcomgzbvg}).  This is consistent with the top equation in~\eq{gcomclbvg}: perturbatively $\check g (M) = - M^2 + \cdots$ which gives (left-hand side) $\check g (0) = 0$ consistent with (right-hand side) $ g ( f (0)) = g (0) = 0$.  Indeed, this requires a rolling dilaton~(\eq{gcomgzbvg}).

Consider now the interior, in particular the AF region $\tilde f\sim f_0$ and 
$\tilde M \sim 0$. Here $\tilde P(f_0) = 0$ (\eq{smlimits}), therefore $\tilde g(f_0)=1$ (\eq{gcomgzbvg}), and there is no need for a rolling dilaton~(\eq{gcomgzbvg}). 
The value $\tilde g(f_0)=1$ 
may sound surprising given that $\tilde M (f_0) =0$ and 
$\tilde{\check g} (\tilde M = 0) = 0$. There is no
contradiction, however, with the bottom equation in~\eq{gcomclbvg}: 
$\tilde{\check g }(\tilde M) =  \tilde M^2 + \cdots$ gives (left-hand side) 
$\tilde {\check g} (\tilde M = 0) = 0$ consistent with (right-hand side) $ \tilde g ( \tilde f (0)) = \tilde g (0) = 0$.  Since both $\tilde M(0)$ and $\tilde M (f_0)$ are zero,
the inverse function $\tilde f (M)$ is necessarily multivalued.  In the domain of definition of the perturbative expansion $\tilde f (0) = 0$, so this expansion
cannot  see what is happening for $\tilde f \sim f_0$.

 The above discussion shows how this regular BH convincingly emerges
 from a suitable choice of $M(f)$.  Still, the  interior solution is approaching 
 a background defined by two-dimensional Minkowski space and a {\em fixed} dilaton.
 Such a background is associated with a $c=2$ conformal field theory, not 
 a $c = 26$ one.  Since classical backgrounds of (bosonic) string field theory are
 $c=26$ conformal field theories, one does not expect the regular BH solution
 to be a string theory solution.  More likely, the choice of $M(f)$ leading to it
 defines $\alpha'$ corrections that do not occur in string theory.   

\subsection{A particular example}

As a proof of the existence of a function with the properties mentioned above, we built a concrete example:
\begin{equation}
h(\xi) = {16 \, \xi_0^3 \, \xi \over (\xi^2 + 3 \xi_0^2)^2} \,. 
\end{equation}
In here $\xi_0$ is a parameter that will be adjusted to make $h$ satisfy all the
requisite conditions. 
The  derivatives of $h$ are
\begin{equation}
h'(\xi) =\  {48\, \xi_0^3 \ (\xi_0^2 - \xi^2) \over (\xi^2 + 3 \xi_0^2)^2} \,, 
\quad h''(\xi) = {192 \, \xi_0^3 \ \xi (\xi^2 - 3 \xi_0^2)\over (\xi^2 + 3 \xi_0^2)^4} \,.
\end{equation}
We see that consistent with conditions~\eq{firstconditionsh} we have $h(0)=0$ and 
finite $h'(0)$.  It is simple to check that, as required by~\eq{small_condition},
$h(\xi) \leq 1$ for all $\xi$, with 
$|h(\pm\xi_0)| = 1$. 
The integral of $h$ is easily calculated
\begin{equation}
\int_{0}^{\xi} d\xi'\, h(\xi') = {8 \xi_0\over 3}  { \xi^2 \over \xi^2 + 3 \xi_0^2} \,.
\end{equation}
This implies that
\begin{equation}
\int_{0}^{\infty} d\xi'\, h(\xi') = {8 \xi_0\over 3} = \alpha < \infty \,,
\end{equation}
so that the finiteness condition expressed in the second relation in~\eq{nodiv_conditions} is obeyed. 
The first and third condition in~\eq{nodiv_conditions} (vanishing of $h$ and $\xi h'$
as $\xi\rightarrow \infty$)   are also satisfied.   Note also that $h$ is an odd function, as required. 

\begin{figure}[h]
	\centering
	\includegraphics[width=0.6\textwidth]{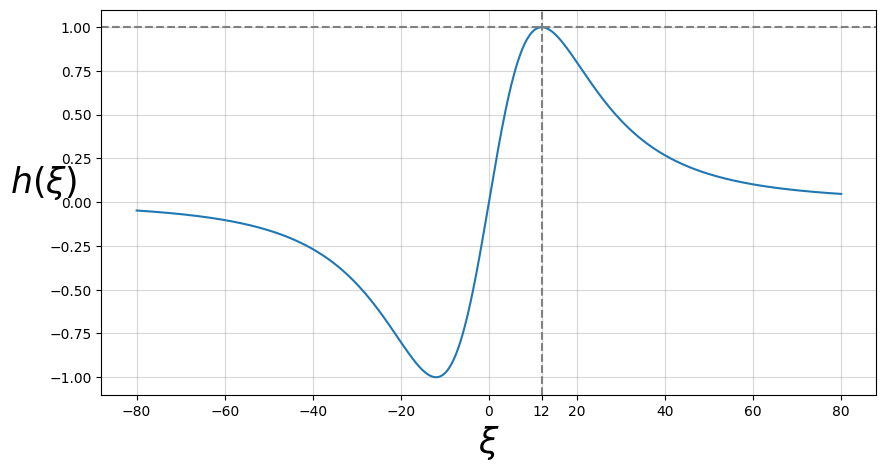}
	\caption{Plot of $h(\xi) = {16 \, \xi_0^3 \, \xi \over (\xi^2 + 3 \xi_0^2)^2}$ with $\xi_0 = 12$. The dotted lines indicate the position of the maximum.}
	\label{fig_hxi_example}
\end{figure}

\begin{figure}[h]
	\centering
\epsfysize=7.0cm
\epsfbox{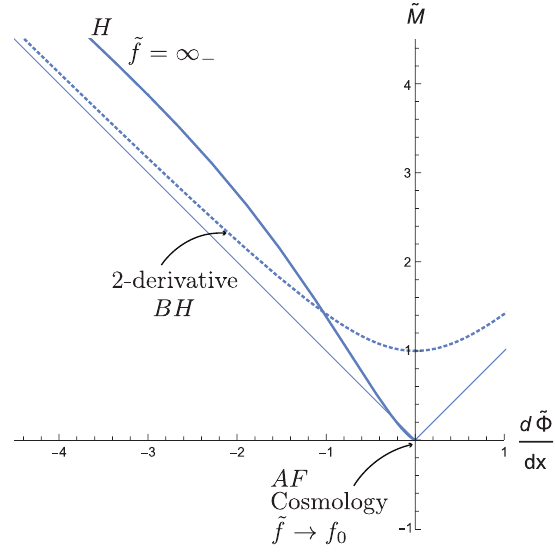}
	\caption{\small Parametric plot for ${d\tilde \Phi\over dx}$ and $\tilde M$ both as functions of $\tilde f\in (12, \infty)$ for the interior of the black hole with resolved singularity (continuous line). Shown dashed is the analogous plot for the two-derivative black hole.  The faint lines are asymptotes at $45^\circ$ and $135^\circ$.  The horizon is far on the $135^\circ$ asymptote, and the late-time asymptotically Minkowski part of the cosmology is around the origin.  }
	\label{figBHext_MP_plots_regular} 
\end{figure}

All of the above conditions were already imposed  in the previous section, but here we have 
the new conditions concerning  the point $\xi_0$, where $h$ reaches the value of one.  
We indeed have, as required by the conditions listed in~\eq{xi0_conditions}, 
\begin{equation}\label{xi0_cnditions}
h(\xi_0) = 1\,, \ \ \ h'(\xi_0) = 0\,, \quad h''(\xi_0) \, = \, 
- \tfrac{3}{2} \tfrac{1}{\xi_0^2} < 0\, . 
\end{equation}
The value of $\xi_0$ is determined by the integral constraint in~\eqref{xi0_conditions} 
\be
\int_{0}^{\xi_0} d\xi'\, h(\xi') = \tfrac{2 \xi_0}{3} =  \xi_0 - 4 \quad \to \quad  
\xi_0 = 12\, .
\ee
We see that for this value of $\xi_0$ we have $\alpha = 32$.  The function $h$ 
has a maximum at $\xi_0$, and being odd, a minimum at $-\xi_0$.  It vanishes
both for $\xi=0$ and as $\xi \to \infty$.
A plot of this function can be seen in Fig.\,\ref{fig_hxi_example}. 
Fig.\,\ref{figBHext_MP_plots_regular} 
shows the plot of $d\tilde\Phi/dx$ and $\tilde M$ as a function of $\tilde f$ for this regular black hole (continuous line).  Superposed in the figure we see the similar plot (Fig.~\ref{bh-interior-nfig}) for the two-derivative black hole
 (dashed lines). 
 The coordinate $x$ for the interior as a function of $\tilde f$ is given by the
 integral
 \be
 \label{xftilrel}
 x \, = \, \int_{\tilde f}^\infty {df\over f \sqrt{\tilde P(f)}} \,, 
 \ee
which correctly gives the horizon $x=0$ for $\tilde f =\infty$ and a $x\to \infty$ for $\tilde f \to f_0$.  The integral above is best done numerically,
and a plot of it is shown in Fig.\,\ref{xfrbh}. 
\begin{figure}[h]
	\centering
\epsfysize=5.0cm
\epsfbox{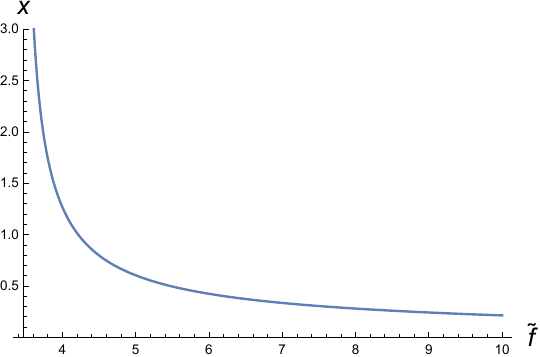}
	\caption{\small The coordinate $x$ as a function of the parameter
	$\tilde f \in (f_0, \infty)$ for the resolved BH interior.  Here $f_0 = \sqrt{12} \sim 3.464$. }
	\label{xfrbh}
\end{figure}

\section{Concluding remarks} \label{conclremks}

In this paper we have considered the theory space of duality-invariant
derivative corrections to the lowest order action that gives a solution
identified as the string theory two-dimensional black hole.  We describe 
the theory space by the choice of function $M(f)$, whose inverse is related
to the usual perturbatively defined $\alpha'$ corrections.  
Using $f$ as a parameter, and with square
roots of nontrivial functions appearing in the equations, $f$ space naturally
becomes a space with branch points and branch cuts.  In this $f$ plane, the 
underlying structure of the solution
is determined by the type of branch points which determine the 
lines or contours that parameterize the solution.  We have argued that
the $f$ parameterization provides an extension of the original $\alpha'$ corrected theory that applies for situations where the series defining the corrections
in the action does not converge.  We have used a notation where $f$ and $\tilde f$
are the parameters for the exterior and interior solutions, respectively.

In fact, we saw that the black hole interior leads to a branch point
at some real, positive $f_0$ with a cut going all the way to $\tilde f=\infty$. 
For the black hole interior with a singularity, the $x$ coordinate reaches
a finite value at the branch point $f_0$, a point that can be reached in finite proper
time.  This means that the space cannot end there, and one must indeed
return to $\tilde f=\infty$ over the cut.  
To avoid the singularity, while preserving a horizon, 
we altered the nature of the branch point $f_0$. Now $x$ reaches
an infinite value as we approach $f_0$, and it takes infinite proper 
time to get there.  Thus we get a complete space without having
to go over the branch.   This yields a black hole  with a horizon but
a regular interior, a regular black hole.  
 The main caveat is that such solution does not
appear to be a string theory one, as discussed in section~\ref{regbhsoljnvg}. 
It could be that while general T-duality invariant
$\alpha'$ corrections allow
for regular black holes, such regular two-dimensional black holes do not arise 
in string theory.  

Our work can be viewed as an exploration of the space 
${\cal T}$ of all possible $\alpha'$ corrections in two dimensions;  the space of
all possible $M(f)$.  If we call  ${\cal S}$ the subspace we showed
leads to conventional singular black holes,  the regular black holes we identified 
appear on some subspace ${\cal R} \subset \partial {\cal S}$  of the boundary of ${\cal S}$. Note that we have not delineated the maximal space $\hat {\cal S} \supset {\cal S}$ that leads to conventional black holes.  Of course, we would very much like to know where $D=2$ string theory sits in ${\cal T}$.  If the string theory black hole
has a conventional singularity, then string theory would lie in $\hat {\cal S}$.

We have also discussed duality in the black hole solutions.
The maximally extended geometry of the black hole of the two-derivative theory
describes a self-dual solution where,  as noted by Giveon, duality maps the horizon
to the curvature singularity.   We showed this is actually a general result.  In particular, 
complete solutions that have horizons but no singularities are not self-dual.
Their T-duals give an alternative description of the space.   

A few issues could be investigated next:
\begin{enumerate}

\item  The present work makes some progress but does not
completely describe the type of situations that can arise throughout
the theory space ${\cal T}$.  
We have seen some
unusual solutions, and it would be of interest to know all that can happen.
We identified a subspace ${\cal R} \subset {\cal T}$ leading to regular black
holes, but there may be a larger subspace $\hat {\cal R}$ giving
non-singular black holes, 
perhaps containing some that could be expected
to be string theory solutions.   Since 2D string theory is strongly believed
to have a black hole, the regions of ${\cal T}$ for which no black hole solution
exists could be thought of as the `swampland' in the theory space.

\item  To better understand solutions, instead of using the $n=1$ gauge, which
results in disconnected $f$ contours for the exterior and interior solutions,
one may try for a better, more geometric understanding in the $nm=1$ gauge,
where maximal extensions may be more easily constructed.  

\item It would  indeed  
be important to determine the maximal extensions of the regular black holes
obtained here
and to display their Penrose diagrams. 
The challenge here is doing the needed work analytically.

\item   
To get hints if the 2D string theory black hole is regular, 
 it may be of interest 
 to do a more careful study of the Dijkgraaf, Verlinde, Verlinde ansatz~\cite{Dijkgraaf:1991ba}  
  to understand its geometry and the behavior of the dilaton. 
In~\cite{Perry:1993ry}, the geometry is first extended to include a 
region with Euclidean signature, which is later removed,  with
completeness restored by gluing infinite number of copies with such excision.
Of course, we still lack 
insight on how this configuration could be a solution of the 
$\alpha'$ corrected {\em string} equations of motion. In~\cite{Codina:2023fhy} we showed that the ansatz, in its given form, does not fit the framework
of our analysis.

\item Our work could help constrain the $\alpha'$ corrections of
string theory.  We have seen that if the string theory BH remains singular then
the $\alpha'$ corrections are strongly constrained.  Indeed for generic
classes of corrections the BH presumably does not exist.  At present the class
of $M(f)$'s for which the BH has no singularity seems strongly restricted, but
further investigation could change this.

\item 
As emphasized in the introduction, our search for a regular BH geometry 
is based on standard general relativity notions, where a singularity 
is probed by point particles and geodesics. Since in string theory  spacetime 
should be probed by strings, the ultimate picture might be quite different. 
As shown here,  the T-dual of a regular BH geometry necessarily features a 
curvature singularity, but it might be that when probed by winding modes 
this singularity is harmless. 
This could be explored with a genuine double field theory \cite{Hull:2009mi,Bonezzi:2023ced} 
on 2D black hole  backgrounds.

\end{enumerate}

\subsection*{Acknowledgements} 

We thank Netta Engelhardt, Maurizio Gasperini, Daniel Grumiller, 
Daniel Harlow, Matthew Kleban,  
Massimo Porrati, Gabriele Veneziano, and Ashoke Sen, for comments and discussions.

This work  is supported by the European Research Council (ERC) under the European Union's Horizon 2020 research and innovation program (grant agreement No 771862). 
T.~C.~is supported by the Deutsche Forschungsgemeinschaft (DFG, German Research Foundation) - Projektnummer 417533893/GRK2575 ``Rethinking Quantum Field Theory".

This material is based upon work supported by the U.S. Department of Energy, Office of Science, Office of High Energy Physics of U.S. Department of Energy under grant Contract Number  DE-SC0012567.(High Energy Theory research).

\appendix

\end{document}